\documentclass[prd,amsmath,floats,amssymb,authoryear,floatfix,superscriptaddress,nofootinbib,onecolumn]{revtex4} 

\usepackage{graphicx}
\usepackage{hhline} 

\usepackage{caption}

\usepackage{ulem}
\usepackage{natbib}
\usepackage{appendix}
\usepackage{multirow}
\usepackage{makecell}
\usepackage{placeins}
\usepackage{mathtools}
\newcommand{\rthis}[1]{\textcolor{black}{#1}}

\renewcommand{\thetable}{\textbf{\arabic{table}}}

\makeatletter
\renewcommand{\fnum@figure}{\textbf{Figure~\thefigure}}
\renewcommand{\fnum@table}{\textbf{Table~\thetable}}
\makeatother

\usepackage{float}
\usepackage[caption=false]{subfig}
\usepackage[paperwidth=210mm,paperheight=297mm,centering,hmargin=1cm,vmargin=2.5cm]{geometry}

\usepackage[plainpages=false, colorlinks=true, anchorcolor=blue, linkcolor=blue, citecolor=blue, bookmarks=false]{hyperref}

\begin{document}
\preprint{APS/123-QED}

\title{Probing cosmic anisotropy with galaxy clusters and supernovae}

\author{Shubham Barua}
 \altaffiliation{Email:ph24resch01006@iith.ac.in}
\author{Sujit K. Dalui}
 \altaffiliation{Email:ph24mscst11036@iith.ac.in}
\author{Shantanu Desai}
 \altaffiliation{Email:shntn05@gmail.com}
\affiliation{
 Department of Physics, IIT Hyderabad Kandi, Telangana 502284,  India}

\begin{abstract}
Using $\Lambda$CDM and Pad\'e-(2,1) cosmography, we study directional variations in the Hubble constant, $H_0$, using galaxy clusters (GC) and Type Ia Supernovae (SNe), from the PantheonPlus compilation, employing the hemisphere decomposition method. Since there is a degeneracy between $H_0$ and absolute magnitude $M_B$ for SNe, Cepheid-host calibration is usually required to constrain $H_0$. To complement the Cepheid-host calibration in the SNe analysis, we use GC X-ray scaling relations as a complementary probe of anisotropy. We consider five redshift bins on the full SNe sample $[z_\mathrm{min}, 0.5]$ where $z_\mathrm{min}= \{0.015, 0.03, 0.05, 0.1, 0.2\}$. Our results show that the combined GC sample is consistent with statistical isotropy and that there is a $\sim 2.5 \sigma$ anisotropy preference in the lowest-redshift bin in the SNe-only case. This preference weakens when low-redshift SNe are progressively removed. In the SNe$+$GC analysis, we find that any anisotropy preference is driven by how the $H_0-M_B$ and the $H_0-k$ (where $k$ is the normalization parameter in the GC scaling relation) degeneracies are treated. The low-redshift bin anisotropy preference in the SNe-only analysis persists in the SNe$+$GC case when Cepheid-host calibration is used while the results are consistent with statistical isotropy when $k$ is fixed. Overall, our results do not provide compelling evidence for a statistically significant anisotropy in the local expansion rate. Our findings highlight the importance of combining complementary cosmological probes and examining the dependence of anisotropy constraints on calibrations and parameter degeneracies when assessing claims of cosmic anisotropy.
\end{abstract}


\maketitle
\section{Introduction}
\label{sec:1}

The cosmological principle~\cite{aluri_2023} requires the universe to be statistically homogeneous and isotropic on sufficiently large scales. The support for the cosmological principle comes mainly from the isotropy of the cosmic microwave background (CMB)~\cite{planck_2020, adams_1998, barriga_2001}. It forms the edifice of modern cosmology. Nevertheless, its validity on large scales has been critically examined through studies of the quasar dipole~\cite{secrest_2021, zhao_2021, hu_2020, guandalin_2023, dam_2023, ferreira_2024}, the radio galaxy dipole~\cite{qiang_2020, singal_2023, wagenveld_2023}, bulk velocity flow~\cite{watkins_2015, watkins_2023, wiltshire_2013, nadolny_2021, courtois_2025}, dark velocity flow~\cite{atrio_2015}, CMB anomalies~\cite{copi_2010, schwarz_2016}, possible SNe dipole and quadrupole~\cite{schwarz_2007, dhawan_2023, cowell_2023, sorrenti_2023, sah_2025, heinesen_2022, javanmardi_2015, salzano_2025, kalus_2013}, Gamma ray bursts (GRBs)~\cite{mondal_2026, andrade_2019, tarnopolski_2017, attila_2019, meegan_1992, dong_2026,Mokal26}, etc. 

Imposing the cosmological principle leads to the  Friedmann-Lemaitre-Robertson-Walker (FLRW) spacetime metric~\cite{weinberg_1972} which forms the background of the highly successful standard $\Lambda$CDM model. Within this framework, the Hubble parameter $H(z)=\frac{\dot{a}}{a}$, where $a(t)$ is the scale factor, describes the expansion rate of the universe at redshift $z$. $H_0 \equiv  H(z=0)$ is the Hubble constant denoting the current expansion rate of the universe. The precise determination of $H_0$ has become a major topic of interest due to the Hubble tension~\cite{ Bethapudi,Verde2019,Shah,Julien,maksym_2025, leandros_2023, capozziello_2024, hu_2023, verde_2024, valentino_2021, skara_2022}, a discrepancy between the early-universe (e.g. CMB) and late-universe (e.g. distance-ladder) measurements. Theoretically, it is expected that $H_0$ must be a constant~\cite{krishnan_2021b} and not depend on the direction or position in space. In principle, if a new degree of freedom associated with anisotropy is introduced into the standard $\Lambda$CDM model, this might potentially alleviate the $H_0$ tension~\cite{tsagas_2010, tsagas_2021, colgain_2019, krishnan_2021, joan_2018, anand_2017, valentino_2017, wang_2021, gomez_2017, valentino_2018}.

Recent studies have suggested that the universe may exhibit departures from large-scale homogeneity and isotropy, including a possible anisotropy in the inferred Hubble parameter values~\cite{koksbang_2021}. Type Ia Supernovae (SNe) have been widely used to test the cosmological principle~\cite{basheer_2023, hu_2024_2, tang_2023, sun_2018, colin_2019, richard_2026, bengaly_2024}. Ref.~\cite{leandros_2023} used the hemisphere comparison method to test the isotropy of the absolute magnitudes ($M_B$) of the PantheonPlus samples in various redshift and distance bins. Their findings indicate sharp changes in the inferred anisotropy at distances less than 40 Mpc. Ref.~\cite{colgain_2023} analyzed the anisotropic distance ladder and found large $H_0$ values in the hemisphere encompassing the CMB dipole direction (see also Ref.~\cite{luongo_2022}). Ref.~\cite{krishnan_2022} emphasizes that the cosmic anisotropy may be due to a breakdown in the cosmological principle or due to statistical fluctuations in the SNe residing in Cepheid-host galaxies. In other works, it has been argued  that the violation of the cosmological principle might be associated with the evolution of $H_0$ with redshift~\cite{malekjani_2024, dainotti_2021, millon_2020}.

The inferred anisotropy can also be affected by the non-uniform angular distribution of SNe across the sky. This can be addressed either by improving the sky coverage with additional independent measurements or by selecting subsamples with uniform angular distributions. In this work, we adopt the former approach and combine galaxy clusters (GC) with the SNe dataset, which provide a complementary probe of anisotropy through their X-ray scaling relations. GC have been used to test cosmic anisotropy extensively~\cite{hu_2026, mikgas_2025}. Combining the two datasets allows us to investigate whether any anisotropy preference found in the SNe sample persists when additional, independent observational constraints are included.

The degeneracy between $H_0$ and $M_B$ is broken by using Cepheid-host SNe, which provide an external constraint on $M_B$. However, it was pointed out in Ref.~\cite{colgain_2023} that directional variations in $M_B$ could lead to variations in the inferred $H_0$. Hence, it is useful to investigate an independent probe that does not rely on the Cepheid distance ladder. In this context, GC scaling relations provide an alternative way of testing directional variations in the inferred expansion rate.

Ref.~\cite{mikgas_2020} stated that one reasonable assumption for GC is that the physics within the intra-cluster medium (ICM) of GC that determines the correlation between the X-ray luminosity $(L_X)$ and temperature $(T)$ should be the same regardless of the direction on the sky. Under this assumption, the  normalization and slope of the $L_X-T$ relation are expected to be isotropic and can be determined from the full GC sample. In our analysis, we therefore consider the case in which normalization $k$ of the $L_X-T$ relation is fixed to its best-fit value obtained under a fiducial cosmology, thereby removing the $H_0-k$ degeneracy and providing a complementary test of the anisotropy inferred from the SNe data.

In this work, we combine SNe and GC  datasets to study the anisotropy in $H_0$ following the methodology of Ref.~\cite{colgain_2023}. For this purpose, we consider the $\Lambda$CDM model and the model-independent Pad\'e-(2,1) cosmography. In Section~\ref{sec2}, we describe the observational data used for this analysis. In Section~\ref{sec3}, we describe our methodology and in Section~\ref{sec4} we present and discuss the results from our analyses. Finally, we summarize our conclusions in Section~\ref{sec5}.

\section{Datasets}
\label{sec2}

\begin{figure*}
    \centering
    \subfloat[GC distribution ($0.004 \leq z \leq 0.447$).\label{1a}]{\includegraphics[width=0.3\textwidth,keepaspectratio]{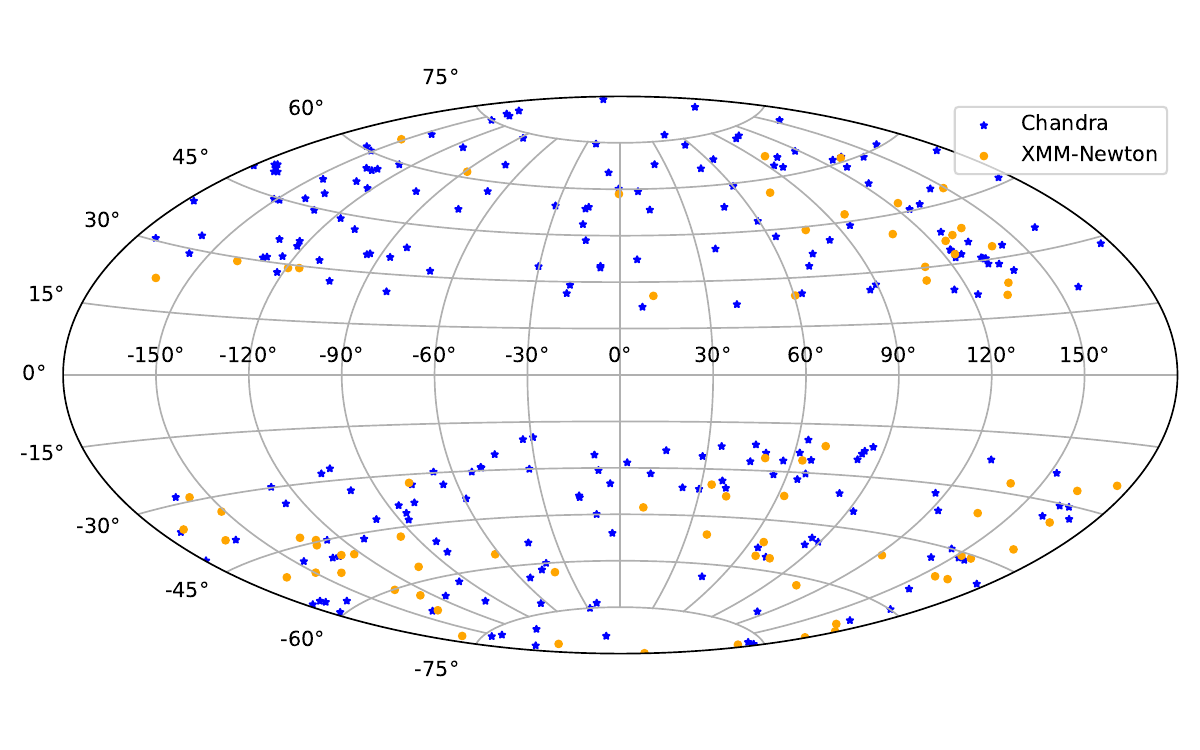}}
    \subfloat[SNe count ($z_\mathrm{min} \leq z \leq 0.5$)\label{1b}]{\includegraphics[width=0.3\textwidth,keepaspectratio]{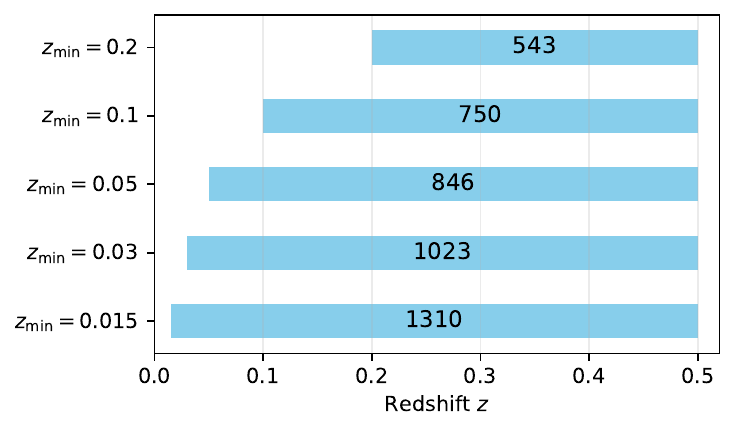}}
    \subfloat[PantheonPlus SNe distribution ($0.015 \leq z \leq 0.5$).\label{1c}]{\includegraphics[width=0.4\textwidth,keepaspectratio]{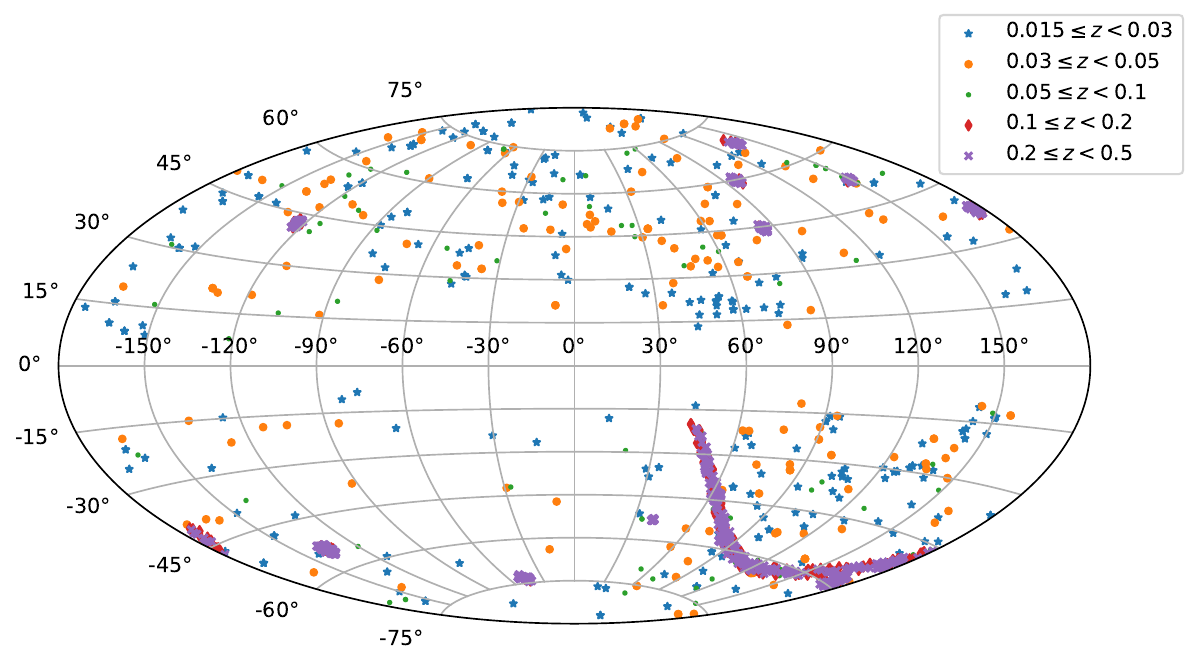}}
    \caption{Distribution of GC (\ref{1a}) and Type Ia SNe (\ref{1c}) in the galactic coordinate system. The 77 Cepheid-host calibrators are not included in the SNe count and distributions shown here except for two which lie within the redshift range $0.015 \leq z \leq 0.5$.}
    \label{fig1}
\end{figure*}

The GC sample used in this work was compiled by Ref.~\cite{mikgas_2020} from the Meta-Catalogue as of July 2019 of X-ray detected Clusters of galaxies [MCXC;~\cite{mcxc_2011}]. The parent catalogs are all based on the ROSAT All-Sky Survey [RASS;~\cite{voges_1999}]. The basic selection criteria used for these clusters are that they lie above a fixed X-ray flux  threshold of $5 \times 10^{-12}$ ergs/s/cm$^2$ in the energy range from 0.1-2.4 keV~\cite{mikgas_2020}.

The combined sample consists of 313 GC in the  redshift range from 0.004 to 0.447. The sample is made up of two datasets: Chandra (237 clusters, $0.004 \leq z \leq 0.447$) and XMM-Newton (76 clusters; $0.018 \leq z \leq 0.244$). The distribution of the GC in the sky is shown in Fig.~\ref{1a}. The spatial distribution of the Chandra dataset is more uniform than that of XMM-Newton. Overall, the GC sample is relatively uniform and is suitable for testing anisotropy.

We utilize Type Ia SNe data from the PantheonPlus (PP) compilation~\cite{scolnic_2022, brout_2022}. It consists of 1701 light curves from 1550 distinct Type Ia SNe and covers a redshift range of 0.001 to 2.26. In Fig.~\ref{1c}, we show the distribution of SNe based on the redshift ranges considered in this work (discussed further in Section~\ref{sec3.2}). The distribution of SNe in the largest redshift interval  $0.015 \leq z \leq 0.5$ is approximately $80\%$ of the total SNe compilation. Hence, as discussed in Section~\ref{sec3}, we consider five redshift intervals $[z_\mathrm{min}, 0.5]$ where $z_\mathrm{min} \in \{0.015, 0.03, 0.05, 0.1, 0.2\}$. In order to use Cepheid-host SNe for calibration, we also consider 77 Cepheid-host SNe in the redshift range $0.00122 \leq z \leq 0.01682$. 

\section{Data Analysis Methods}
\label{sec3}

The physical quantities of GC follow tight scaling relations~\cite{kaiser_1986}. Specifically, the correlation between $L_X$ and $T$ of GC is of particular interest in cosmology. The general properties of the $L_X-T$ scaling relation have been extensively studied  in the literature~\cite{mikgas_2020, mikgas_2018, mikgas_2021, vikhlinin_2002, pacaud_2007, pratt_2009, mittal_2011, reichert_2011, hilton_2012, maughan_2012, bharadwaj_2015, lovisari_2015, giles_2016, zou_2016}. The $L_X-T$ relation is of the form~\cite{mikgas_2020}
\begin{equation}
    \label{eqn1}
    \frac{L_X}{10^{44}\mathrm{erg/s}}E(z)^{-1} = k \left(\frac{T}{4\mathrm{keV}}\right)^s,
\end{equation}
where $E(z) = H(z)/H_0$ scales $L_X$ accordingly to explain the redshift evolution of the $L_X-T$ relation. The parameters $k$ and $s$ are the normalization and slope of the scaling-relation, respectively. $L_X$ can be derived from the observed $k$-corrected flux $F$ using $L_X=4\pi d_L^2F$ where $d_L$ is the luminosity distance~\cite{mikgas_2020}. 
It is given by
\begin{equation}
    \label{eqn2}
    d_L = \frac{c(1+z)}{H_0}\int_0^z\frac{dz'}{E(z')}.
\end{equation}
Eq.~\ref{eqn1} can be written in logarithmic form as 
\begin{equation}
    \label{eqn3}
    \log L_X' = \log k + s\log \tau,
\end{equation}
where 
\begin{equation}
    \label{eqn4}
    L_X' = \frac{L_X}{10^{44}\mathrm{erg/s}}E(z)^{-1} \mathrm{\ and} \ \tau=\frac{T}{4\mathrm{keV}}.
\end{equation}
The corresponding $\chi^2$ expression involving uncertainties in both $x$ and $y$ variable  is given by~\cite{gauri_2024}:
\begin{equation}
    \label{eqn5}
    \chi^2_\mathrm{cluster}=\sum_{i=1}^N\left(\log(2\pi\sigma^2_\mathrm{tot})+\frac{(\log[L'_{X\mathrm{,obs}}]-\log[L'_{X\mathrm{,th}}(\tau, \boldsymbol{\theta})])^2}{\sigma_\mathrm{tot}^2}\right),
\end{equation}
where $\sigma_\mathrm{tot}^2 = \sigma^2_{\log L_i}+s^2\sigma^2_{\log T_i}+\sigma^2_\mathrm{int}$, $N$ is the number of clusters, $L'_{X\mathrm{,th}}$ is the theoretical X-ray luminosity, $L'_{X\mathrm{,obs}}$ is the observed X-ray luminosity (which we compute from $F$ in this work, as mentioned above); $\boldsymbol{\theta}$ represents the parameters to be fitted; $\sigma_{\log L_i}$ and $\sigma_{\log T_i}$ are the $1\sigma$ errors for luminosity and temperature, respectively~\footnote{$\sigma_{\log x}=\log(e)\frac{x^++x^-}{2x}$ where $x^+$ and $x^-$ \rthis{denote} the upper and lower $1\sigma$ uncertainties of a quantity $x$, \rthis{respectively} (Ref.~\cite{mikgas_2020}).}, while $\sigma_\mathrm{int}$ is the intrinsic scatter of the $L_X-T$ correlation. 

The $\chi^2$ expression for SNe is given by 
\begin{equation}
    \label{eqn6}
    \chi^2_{\mathrm{SNe}} = \vec{Q}^T \cdot C_{\mathrm{stat+sys}}^{-1} \cdot \vec{Q},
\end{equation}
where $C$ denotes the PP covariance matrix and the vector $\vec{Q}$ is defined as:
\begin{equation}
    \label{eqn7}
    Q_i=\begin{cases}
            m_i - M_B - \mu_{\mathrm{ceph},i} \ , & \ i \in \mathrm{Cepheid\ hosts,} \\
            m_i - M_B - \mu_\mathrm{model}(z_i) \ , & \ \mathrm{otherwise,}
        \end{cases}
\end{equation}
where $m_i$ is the apparent magnitude of the SNe, $M_B$ is the peak absolute magnitude and $\mu_\mathrm{model}$ denotes the distance modulus given by 
\begin{equation}
    \label{eqn8}
    \mu_\mathrm{model}(z) = m-M_B = 5\log\left[\frac{d_L(z, \boldsymbol{\theta})}{\mathrm{Mpc}}\right]+25,
\end{equation}
where $d_L(z)$ is given by Eq.~\ref{eqn2} and it depends on the cosmological parameters, including $H_0$. Eq.\ref{eqn7} is used when using Cepheid-hosts to break the $H_0-M_B$ degeneracy. When using uncalibrated SNe, the vector $Q$ is defined as $m_i - M_B - \mu_\mathrm{model}(z_i)$.

In this work we consider two cosmological models. The first is the standard $\Lambda$CDM model given by
\begin{equation}
    \label{eqn9}
    H(z) = H_0\sqrt{\Omega_m(1+z)^3 + 1 - \Omega_m},
\end{equation}
where $\Omega_m$ denotes the matter density of the universe at $z = 0$. We also consider the Pad\'e-(2,1) cosmography~\cite{visser_2015, capozziello_2018, lusso_2019, bargiacchi_2021, hu_2024} for which the Hubble parameter is given by 
\begin{equation}
    \label{eqn10}
    H(z) = H_0\frac{\left[2(1+z)^2(3+z+j_0z-q_0(3+z+3q_0z))^2\right]}{p_0+p_1z+p_2z^2}, 
\end{equation}
where $p_0 = 18(q_0-1)^2$, $p_1=6(q_0-1)(-5-2j_0+8q_0+3q_0^2)$ and $p_2=14+7j_0+2j_0^2-10(4+j_0)q_0+(17-9j_0)q_0^2+18q_0^3+9q_0^4$. The analytical expression for the luminosity distance in the case of Pad\'e-(2,1) cosmography is given by
\begin{equation}
    \label{eqn11}
    d_L = \frac{c}{H_0}\left[\frac{z(6(q_0-1)+(-5-2j_0+q_0(8+3q_0))z)}{-2(3+z+j_0z)+2q_0(3+z+3q_0z)}\right].
\end{equation}
Throughout this work, we fix $j_0$ to one.

\subsection{Anisotropy analysis}
\label{sec3.1}

We follow the methodology of~\cite{colgain_2023} in order to test the directional variations in $H_0$. The right ascension and declination angles on the sky are first converted to galactic coordinates $(l, b)$ since SNe positions are in equatorial coordinates while GC positions are in galactic coordinates. Next, we construct vectors using the identity
\begin{equation}
    \label{eqn12}
    \vec{v} = (\mathrm{cos}\ b\ \mathrm{cos}\ l, \mathrm{cos}\ b\ \mathrm{sin}\ l, \mathrm{sin}\ b).
\end{equation}
We first construct a grid of points on the sky $(l, b)$ using \texttt{HEALPix}~\cite{healpix} with N$_\mathrm{side}$ = 8 (corresponding to 768 equal-area pixels). For each grid point, we compute the corresponding unit direction vector on the sky, $\vec{v}_\mathrm{sky}$. Similarly, for each GC and SNe, we construct the unit vectors $\vec{v}_\mathrm{gc,i}$ and $\vec{v}_\mathrm{sne,i}$. Depending on the inner products $\vec{v}_\mathrm{sky}\cdot \vec{v}_\mathrm{gc,i}$ and $\vec{v}_\mathrm{sky}\cdot \vec{v}_\mathrm{sne,i}$, we separate the SNe and GC into two ($+$ and $-$) hemispheres. The likelihoods for each hemisphere are constructed according to Equations~\ref{eqn5}, \ref{eqn6}, and~\ref{eqn7} depending on which dataset combination is used. 
We then extremize the likelihood to find the best-fit values of parameter combinations from the list $(H_0, \Omega_m, q_0, M_B,k,s,\sigma_\mathrm{int})$ in each hemisphere for a particular sky grid point.  The exact parameter combination varies \rthis{depending} on the  dataset combination, as discussed in Section~\ref{sec3.2}. The optimization is performed using the \texttt{Minuit} minimizer~\cite{minuit} from the \texttt{iminuit} \footnote{\url{https://zenodo.org/records/17565861}} package. The $1\sigma$ parameter uncertainties are estimated using the \texttt{Hesse} method of \texttt{iminuit} which computes the inverse of the Hessian matrix of the $\chi^2$ function at the minimum~\cite{leandros_2023_2, colgain_2023}. 

In the SNe $\chi^2$ expression (Eqn.~\ref{eqn6}), we retain the full covariance structure between the two hemispheres. For a given sky direction, the SNe sample is divided into the $+$ and $-$ hemispheres (as discussed above) and the corresponding residual vectors are denoted as $\vec{Q}^+$ and $\vec{Q}^-$. These are then combined into a single residual vector,
$\begin{pmatrix}
    \vec{Q}^+ \\
    \vec{Q}^-
\end{pmatrix}$.

The corresponding covariance matrix $C$ is constructed from the full PP covariance matrix after selecting all SNe belonging to either hemisphere,
$\begin{pmatrix}
    C^{++} & C^{+-} \\
    C^{-+} & C^{--}
\end{pmatrix}$, where $C^{++}$ and $C^{--}$ represent the covariance submatrices within the two hemispheres, while $C^{+-}$ and $C^{-+}$ encode the correlations between SNe located in opposite hemispheres. This treatment preserves all the  covariance information present in the PP dataset including correlations between SNe belonging to different hemispheres. By retaining the full covariance matrix, our analysis provides a statistically consistent assessment of hemispherical anisotropy.

Following Ref.~\cite{hu_2024}, we compute the anisotropy level which describes the degree of deviation from isotropy and is given by
\begin{equation}
    \label{eqn13}
    AL(\theta) = 2\left(\frac{\theta^+-\theta^-}{\theta^+ +\theta^-}\right), 
\end{equation}
where $\theta$ is the parameter of interest (e.g., $H_0$) and the $+$ and $-$ refer to the northern and southern hemispheres. The corresponding uncertainty is given by  \footnote{This is the full uncertainty obtained by using error propagation of Eq.\ref{eqn13} unlike Ref.~\cite{hu_2024}.}
\begin{equation}
    \label{eqn14}
    \delta_{AL}^{\theta} = \frac{4}{(\theta^++\theta^-)^2}\sqrt{(\delta\theta^{+})^2(\theta^{-})^2+(\delta\theta^{-})^2(\theta^{+})^2}.
\end{equation}
where $\delta\theta$ denotes the uncertainty in $\theta$.

To assess whether the observed anisotropy is statistically significant, we perform a Monte Carlo test considering 1000 mock realizations. For each realization, we keep the sky positions of the real sample - $(l, b)$ of each SNe (and/ or GC)  unchanged - while randomly reshuffling the corresponding measured quantities ($z$, apparent magnitudes, Cepheid-host distance modulus, calibrator flag for SN; $z$, luminosity, flux, and temperature-related quantities for GC). The covariance matrix for SNe is not regenerated for each realization; rather, its rows and columns are permuted using the same random ordering applied to the SNe data, thereby preserving the covariance structure among the reshuffled SNe quantities. This method preserves the true angular directions of the real sample while destroying any genuine correlation between the sky position and physical measurements, thereby generating a realization which is consistent with statistical isotropy.

For each of the 1000 reshuffled realizations, we repeat the full hemispherical scan (described above) and record the maximum anisotropy level, $AL_\mathrm{mock}^\mathrm{max}$. The global $p$-value ($\mathrm{p}_\mathrm{full}$) is then estimated as the fraction of mock realizations for which $AL_\mathrm{mock}^\mathrm{max} \geq AL_\mathrm{obs}^\mathrm{max}$. We define our null hypothesis as the assumption that the sample is statistically isotropic, i.e. any variations in the best-fit parameters arise purely from random sampling and measurement noise, with no dependence on sky position. $\mathrm{p}_\mathrm{full}$ quantifies the probability of obtaining an anisotropy signal ($AL$) at least as large as the observed one under this null hypothesis.

To visualize the directional variations across the sky, we interpolate the values of $AL(\theta)$ computed at each of the scanned {\tt HEALPix} grid points onto a dense regularly-spaced angular mesh, using cubic interpolation via the $\texttt{Python}$ \texttt{scipy} library (\texttt{scipy.interpolate.griddata}). The reported maximum values of $AL(\theta)$, along with its corresponding sky directions, are always identified directly from the discrete set of scanned grid points, without any reliance on the interpolated map.

\subsection{Parameter and dataset combinations}
\label{sec3.2}

There are two sets of parameters involved: the $L_X-T$ scaling-relation parameters \{$k$, $s$, $\sigma_\mathrm{int}$\} for the GC data and cosmological parameters \{$H_0, \Omega_m, q_0, M_B$\} ($M_B$ is the Cepheid-host calibration parameter from SNe).

Before performing anisotropy analysis, the scaling relation parameters $k, s$, and $\sigma_\mathrm{int}$ are determined once, globally, for each of the three GC samples (XMM-Newton, Chandra and their combination), under both $\Lambda$CDM and Pad\'e-(2,1) cosmography. For this step, the cosmological parameters are fixed to their fiducial values, $H_0 = 70$~km/s/Mpc, $\Omega_m = 0.3$ (for $\Lambda$CDM) or $q_0 = -0.55$ (for cosmography), while \{$k, s, \sigma_\mathrm{int}$\} are constrained using MCMC sampling (\texttt{Cobaya} with the \texttt{BOBYQA} minimizer~\cite{bobqa_2018, bobqa_2018_2} and \texttt{Getdist}~\cite{getdist_2025} for posterior analysis). A direct likelihood maximization using \texttt{iminuit} gives comparable best-fit values. \rthis{This fiducial cosmology is used to obtain the best-fit values of \{$k, s, \sigma_\mathrm{int}$\}. In the GC-only analysis, we fix $\Omega_m =0.3 \ (q_0=-0.55)$, consistently with the fiducial cosmology adopted to determine $k$. We note that allowing $\Omega_m$ to vary does not significantly affect the inferred $AL(H_0)$, as low-$z$ GC weakly constrain $\Omega_m$.}

We consider three dataset combinations: GC, SNe and SNe$+$GC. For the SNe$+$GC combination, $H_0$ is degenerate with both the $L_X-T$ normalization parameter $k$ and $M_B$. We therefore consider two cases: 
\begin{itemize}
\item $k$ is fixed to its fiducial best-fit value, removing the $H_0-k$ degeneracy; or 
\item  $k$ is treated as a free parameter and Cepheid-host SNe are used to calibrate $M_B$, thereby removing the $H_0-M_B$ degeneracy. 
\end{itemize}
Table~\ref{table1} summarizes the four resulting parameter combinations in the anisotropy analysis.

For the SNe only and SNe$+$GC analyses, we progressively exclude SNe from the low-redshift end of the sample to assess whether the inferred anisotropy is sensitive to the inclusion of nearby SNe where local effects may be important. We consider the redshift intervals $[z_\mathrm{min}, 0.5]$ where $z_\mathrm{min}=\{0.015, 0.03, 0.05, 0.1, 0.2\}$. In addition, for all our analysis, we consider all 77 Cepheid-hosts in the redshift range $0.00122 \leq z \leq 0.01682$ irrespective of the redshift cut considered or whether we fix $k$ or calibrate the SNe sample using Cepheid-hosts. For each redshift bin (SNe$+$GC analyses), the full GC sample (313) is retained, allowing us to isolate the dependence of the inferred anisotropy on the SNe redshift range while keeping the GC contribution unchanged. We consider the lower redshift threshold $z_\mathrm{min}=0.015$ motivated by the fact that peculiar velocities and nonlinear local structures can significantly affect observations~\cite{lopes_2024, peterson_2022}.

\begin{table}[ht]
\centering
\caption{\rthis{Summary of dataset combinations and free parameters used in the anisotropy analysis. In all cases, $H_0$ is free, with $j_0$ fixed to 1 throughout. For the GC-only analysis, $\Omega_m = 0.3$ ($q_0 = -0.55$) is fixed. $k$, $s$, and $\sigma_\mathrm{int}$ denote the $L_X$--$T$ scaling-relation parameters, and $M_B$ denotes the SNe peak absolute magnitude calibration parameter.}}
\label{table1}
\begin{tabular}{c@{\hspace{0.5cm}}|@{\hspace{0.5cm}}c@{\hspace{0.5cm}}|@{\hspace{0.5cm}}c@{\hspace{0.5cm}}|@{\hspace{0.5cm}}c@{\hspace{0.5cm}}}
\hline
Dataset & $k$ & $M_B$ & Free parameters \\
\hline
GC only & fixed & --- & \{$H_0, s, \sigma_\mathrm{int}$\} \\
SNe only & --- & Cepheid-calibrated & \{$H_0, \Omega_m~(q_0), M_B$\} \\
SNe$+$GC & fixed & --- & \{$H_0, \Omega_m~(q_0), s, \sigma_\mathrm{int}, M_B$\} \\
SNe$+$GC & free & Cepheid-calibrated & \{$H_0, \Omega_m~(q_0), k, s, \sigma_\mathrm{int}, M_B$\} \\
\hline
\end{tabular}
\end{table}

\section{Results and Discussion}
\label{sec4}

In Fig.~\ref{fig1}, we show the distribution of GC and SNe in the Galactic coordinate system. We notice that GC are homogeneously distributed, while for the PP SNe compilation the distribution is not homogeneous above redshift of 0.1, leading to a belt-like structure.

\begin{figure}[ht]
    \centering
    \subfloat[$\Lambda$CDM\label{2a}]{\includegraphics[width=0.5\textwidth,keepaspectratio]{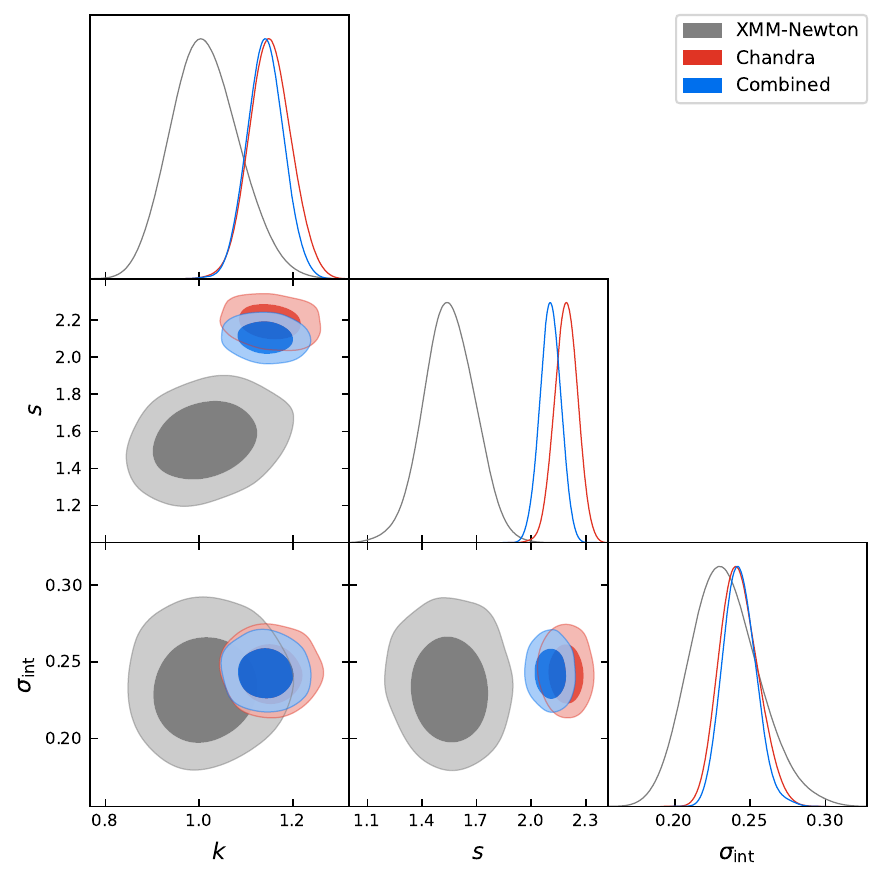}}
    \subfloat[Pad\'e-(2,1) cosmography\label{2b}]{\includegraphics[width=0.5\textwidth,keepaspectratio]{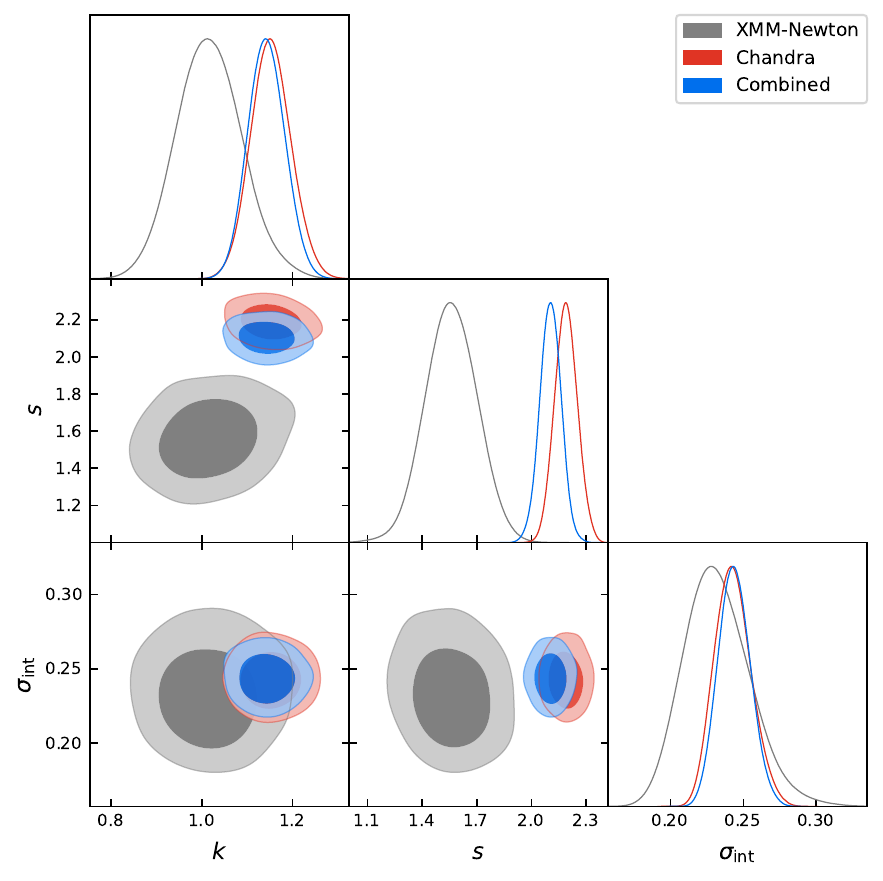}}
    \caption{68\% confidence contours of the $L_X-T$ scaling-relation parameters ($k$, $s$, $\sigma_\mathrm{int}$) for different datasets, assuming $\Lambda$CDM (\ref{2a}) and Pad\'e-(2,1) cosmography (\ref{2b}).}
    \label{fig2}
\end{figure}

We first constrain the $L_X-T$ scaling-relation parameters in order to proceed with the anisotropy analyses for datasets involving GC. For this we consider the fiducial standard $\Lambda$CDM cosmology. We find that the values of $k$, $s$, and $\sigma_\mathrm{int}$ are similar for both $\Lambda$CDM and the Pad\'e-(2,1) cosmography for each of the three GC datasets (Chandra, XMM-Newton, and Chandra$+$XMM-Newton). The results can be seen in Fig.~\ref{fig2} and Table~\ref{table2}. We also determine the best-fit values of the scaling-relation parameters which we later use when fixing the parameters.

\begin{table}[ht]
\caption{Scaling-relation parameters ($k$, $s$, $\sigma_\mathrm{int}$) constraints. Values in brackets are the best-fit values.}
\label{table2}
\centering
    \begin{tabular}{c@{\hspace{1cm}}c@{\hspace{1cm}}c@{\hspace{1cm}}c@{\hspace{1cm}}c}
    \hline
    \thead{Dataset} & \thead{$k$} & \thead{$s$} & \thead{$\sigma_\mathrm{int}$}\\
    \hline
    \hline
    & & & \\[0.5ex] 
    \textbf{$\Lambda$CDM} & & & \\[1.5ex]
    Chandra      &   $1.152\pm0.045\ (1.15)$ & $2.192\pm0.063\ (2.19)$ & $0.243^{+0.011}_{-0.014}\ (0.24)$\\[1ex]
    XMM-Newton   &   $1.013^{+0.067}_{-0.078}\ (1.01)$ & $1.55\pm0.14\ (1.55)$ & $0.233^{+0.020}_{-0.025}\ (0.24)$\\[1ex]
    Combined     &   $1.141\pm0.039\ (1.14)$ & $2.106\pm0.057\ (2.11)$ & $0.243^{+0.010}_{-0.012}\ (0.24)$\\[1ex]
    & & & \\[0.5ex]
    \textbf{Pad\'e-(2,1) Cosmography} & & & \\[1.5ex]
    Chandra       &   $1.152\pm0.044\ (1.15)$ & $2.189\pm0.062\ (2.19)$ & $0.243^{+0.011}_{-0.014}\ (0.24)$\\[1ex]
    XMM-Newton    &   $1.016\pm0.072\ (1.01)$ & $1.56\pm0.14\ (1.55)$ & $0.232^{+0.020}_{-0.025}\ (0.24)$\\[1ex]
    Combined      &   $1.143\pm0.040\ (1.14)$ & $2.104\pm0.058\ (2.11)$ & $0.244\pm0.011\ (0.24)$\\[1ex]
    & & & \\[0.5ex]
    \hline
    \end{tabular}
\end{table}

\begin{figure}[ht]
    \centering
    \subfloat[$\Lambda$CDM: XMM-Newton ($k = 1.01$)\label{3a}]{\includegraphics[width=0.33\textwidth,keepaspectratio]{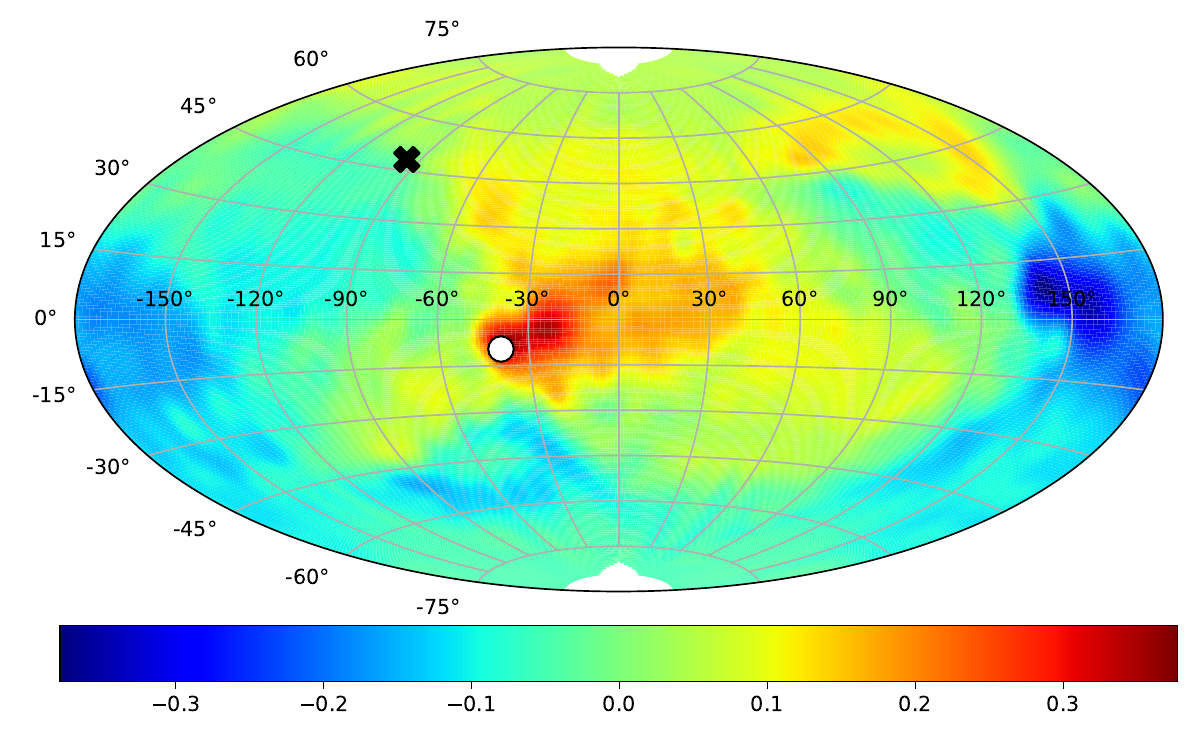}}
    \subfloat[$\Lambda$CDM: Chandra ($k = 1.15$)\label{3b}]{\includegraphics[width=0.33\textwidth,keepaspectratio]{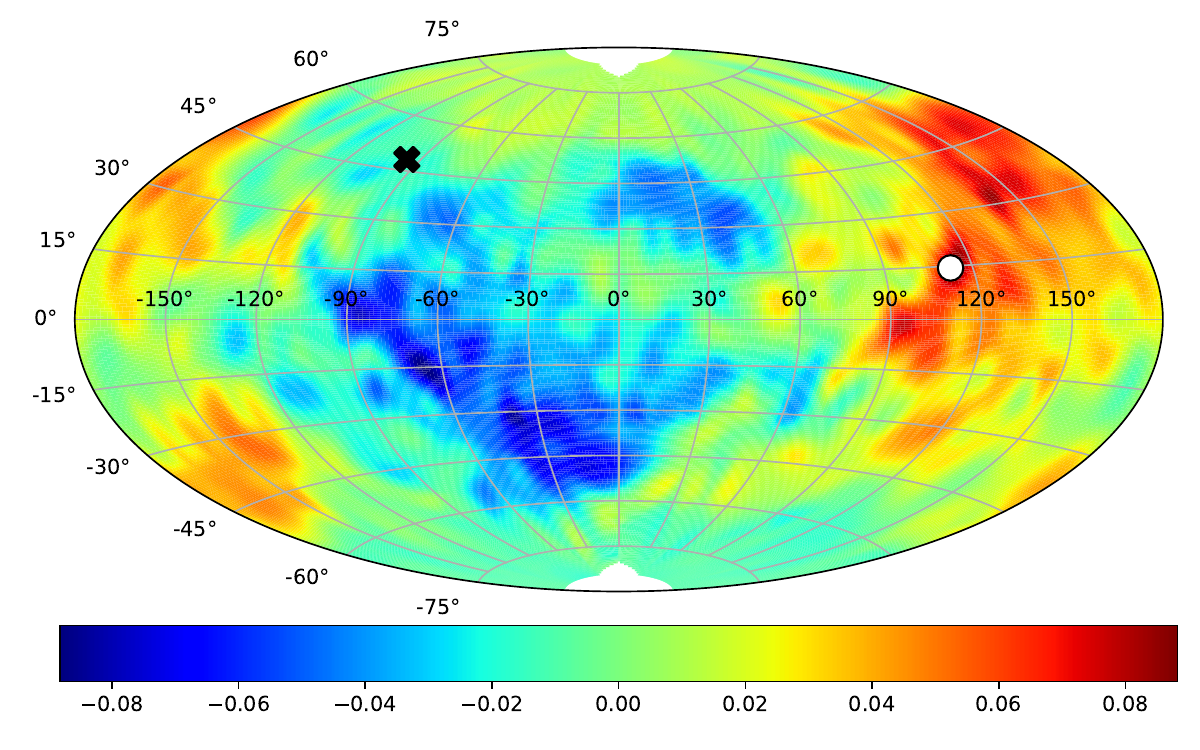}}
    \subfloat[$\Lambda$CDM: Combined sample ($k = 1.14$)\label{3c}]{\includegraphics[width=0.33\textwidth,keepaspectratio]{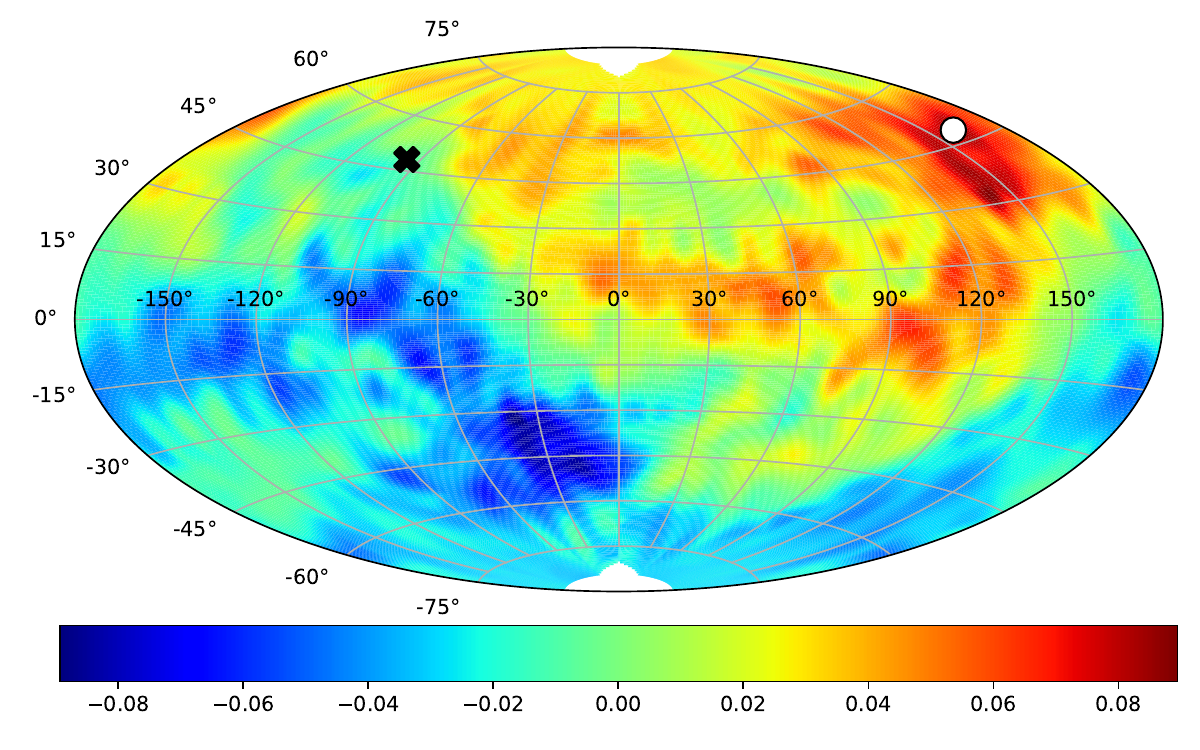}}\\
    \subfloat[XMM-Newton\label{3d}]{\includegraphics[width=0.33\textwidth,keepaspectratio]{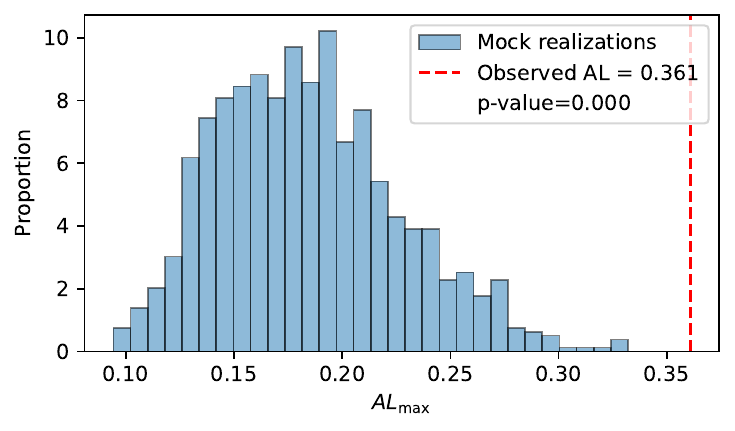}}
    \subfloat[Chandra\label{3e}]{\includegraphics[width=0.33\textwidth,keepaspectratio]{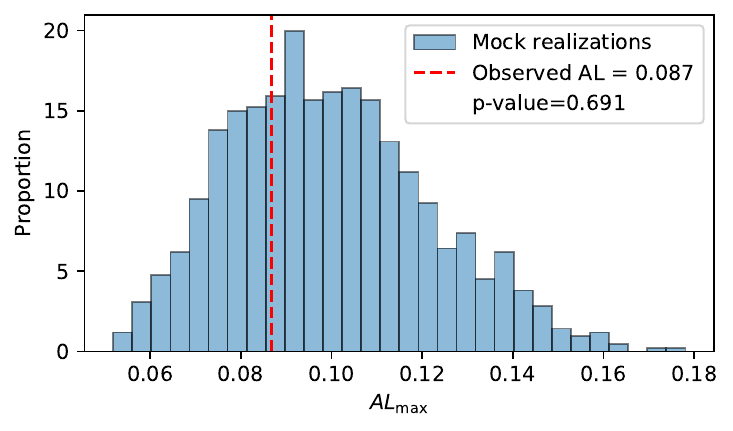}}
    \subfloat[Combined GC\label{3f}]{\includegraphics[width=0.33\textwidth,keepaspectratio]{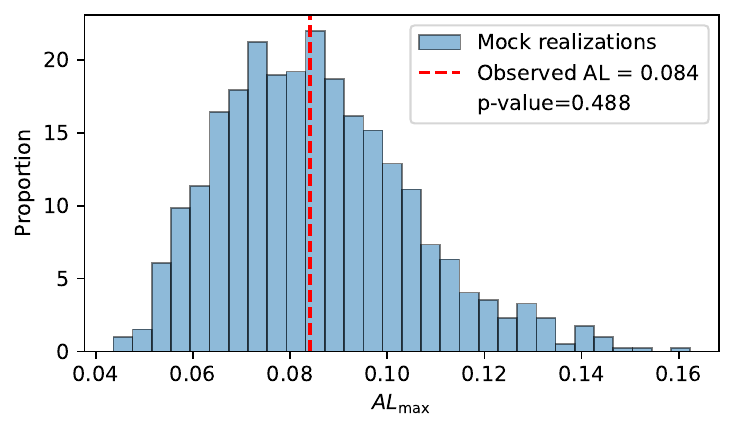}}
    \caption{Sky maps of $AL$ (first row) and the corresponding histograms of the maximum $AL$ values obtained from mock realizations (second row) for the XMM-Newton, Chandra and the Combined dataset assuming $\Lambda$CDM (\rthis{$\Omega_m = 0.3$ is fixed}). The black cross marks the direction of the CMB dipole while the white circle indicates the location of maximum $AL$. The colors correspond to the value of $AL$. For each case, the value to which $k$ is fixed is also stated.}
    \label{fig3}
\end{figure}

In Table~\ref{table3}, we report the maximum $AL$ values, their directions, and the $p$-values ($\mathrm{p}_\mathrm{full}$) obtained from 1000 mock simulations for both $\Lambda$CDM and Pad\'e-(2,1) cosmography. \rthis{We also report the maximum normalized statistic, max($\sigma$) = max[$AL(H_0)/\delta_{AL}^{H_0}$], and its associated $p$-value, $\mathrm{p}_\sigma$, estimated from the mock distribution of the maximum normalized statistic. While $\mathrm{p}_\mathrm{full}$ indicates how unusual the maximum $AL(H_0)$ is in the mock simulations, $\mathrm{p}_\sigma$ indicates how unusual the maximum normalized statistic is, while accounting for the uncertainties in the directional estimates of $H_0$.} We show the $AL$ sky maps and the corresponding maximum $AL$ histograms obtained from the mock realizations in Fig.~\ref{fig3}. \rthis{We find that, for each dataset, both the direction as well as the maximum $AL$ value are similar for $\Lambda$CDM and Pad\'e-(2,1) cosmography.} We therefore conclude that, for GC, the choice between $\Lambda$CDM and cosmography has negligible effect on the inferred anisotropy.

\rthis{We note that the maximum $AL(H_0)$ value for XMM-Newton differs from those for Chandra and the combined sample by $\sim 2\sigma$, and its $\mathrm{p}_\mathrm{full}$, obtained from mock simulations, is formally low.} In contrast, Chandra and the combined sample are both consistent with isotropy, under both $\mathrm{p}_{\sigma}$ and $\mathrm{p}_\mathrm{full}$. Furthermore, the maximum $AL$ direction for XMM-Newton lies at an angular separation of $\sim 138^\circ$ from that of the combined sample. \rthis{The maximum $AL(H_0)$ for XMM-Newton comes from a split in which one hemisphere contains only 13 clusters. Since $AL(H_0)$ is not normalized by its uncertainty, it can be dominated by such poorly constrained splits. After normalizing, we find max($\sigma$) = $3.43$ and its corresponding $\mathrm{p}_{\sigma} \sim 0.16$ \footnote{\rthis{The directions of max($\sigma$) and maximum $AL(H_0)$ need not coincide. For XMM-Newton, the separation between them is $\sim 17^\circ$ whereas for Chandra and the combined sample, they coincide.}}. Although the maximum normalized statistic is large, it is not unusual compared with the isotropic mock realizations, suggesting that the apparent anisotropy may arise from statistical fluctuations in poorly constrained directions}. This interpretation is also consistent with Refs.~\cite{mikgas_2021, He_2025}, who note that the $L_X-T$ scaling relation is a comparatively noisy tracer of anisotropy for this cluster sample: Ref.~\cite{mikgas_2021} found only a mild ($2.4\sigma$) anisotropic signal using the $L_X-T$, with their strongest constraint instead coming from the $Y_\mathrm{SZ}-T$ relation. Ref.~\cite{He_2025} found that anisotropy amplitudes are highly sensitive to the intrinsic scatter in scaling relations, with tighter relations providing stronger constraints. We note that both studies were conducted on the combined sample rather than XMM-Newton alone.

\begin{table}[ht]
\caption{\rthis{Maximum $AL(H_0)$ (Eq.~\ref{eqn13}), its uncertainty $\delta_{AL}^{H_0}$ (Eq.~\ref{eqn14}), the galactic coordinates $(l,b)$ corresponding to the maximum $AL(H_0)$, the maximum normalized statistic max($\sigma$) = max($AL(H_0)/\delta_{AL}^{H_0}$), $\mathrm{p}_\mathrm{\sigma}$ estimated from mock realizations of the maximum normalized statistic, and the global anisotropy $p$-values ($\mathrm{p}_\mathrm{full}$) estimated from mock realizations for different GC datasets.}}
\label{table3}
\centering
    \begin{tabular}{c@{\hspace{1cm}}c@{\hspace{1cm}}c@{\hspace{1cm}}c@{\hspace{1cm}}c@{\hspace{1cm}}c@{\hspace{1cm}}c}
    \hline
    \thead{Dataset} & \thead{$AL(H_0)$} & \thead{$\delta_{AL}^{H_0}$} &\thead{$(l,b)$} & \thead{max($\sigma$)} & \thead{$\mathrm{p}_\mathrm{\sigma}$} & \thead{$\mathrm{p}_\mathrm{full}$}\\
    \hline
    \hline
    & & & & & & \\[0.5ex]
    \textbf{$\Lambda$CDM} & & & & & & \\[1.5ex]
    XMM-Newton & $0.361$ & $0.130$ & $(320.62^\circ, -9.59^\circ)$ & $3.43$ & $0.161$ & $<0.001$ \\
    Chandra    & $0.087$ & $0.038$ & $(112.50^\circ, 14.48^\circ)$ & $2.29$ & $0.675$ & $0.691$ \\
    Combined   & $0.084$ & $0.034$ & $(160.71^\circ, 48.14^\circ)$ & $2.51$ & $0.463$ & $0.488$ \\
    & & & & & & \\[0.5ex]
    \textbf{Pad\'e-(2,1) Cosmography} & & & & & & \\[1.5ex]
    XMM-Newton & $0.361$ & $0.130$ & $(320.62^\circ, -9.59^\circ)$ & $3.43$ & $0.161$ & $<0.001$\\
    Chandra    & $0.087$ & $0.038$ & $(112.50^\circ, 14.48^\circ)$ & $2.29$ & $0.677$ & $0.697$ \\
    Combined   & $0.084$ & $0.034$ & $(160.71^\circ, 48.14^\circ)$ & $2.51$ & $0.462$ & $0.483$ \\
    & & & & & & \\[0.5ex]
    \hline
    \end{tabular}
\end{table}

\begin{figure}[ht]
    \centering
    \subfloat[$0.015 \leq z \leq 0.5$\label{4a}]{\includegraphics[width=0.25\textwidth,keepaspectratio]{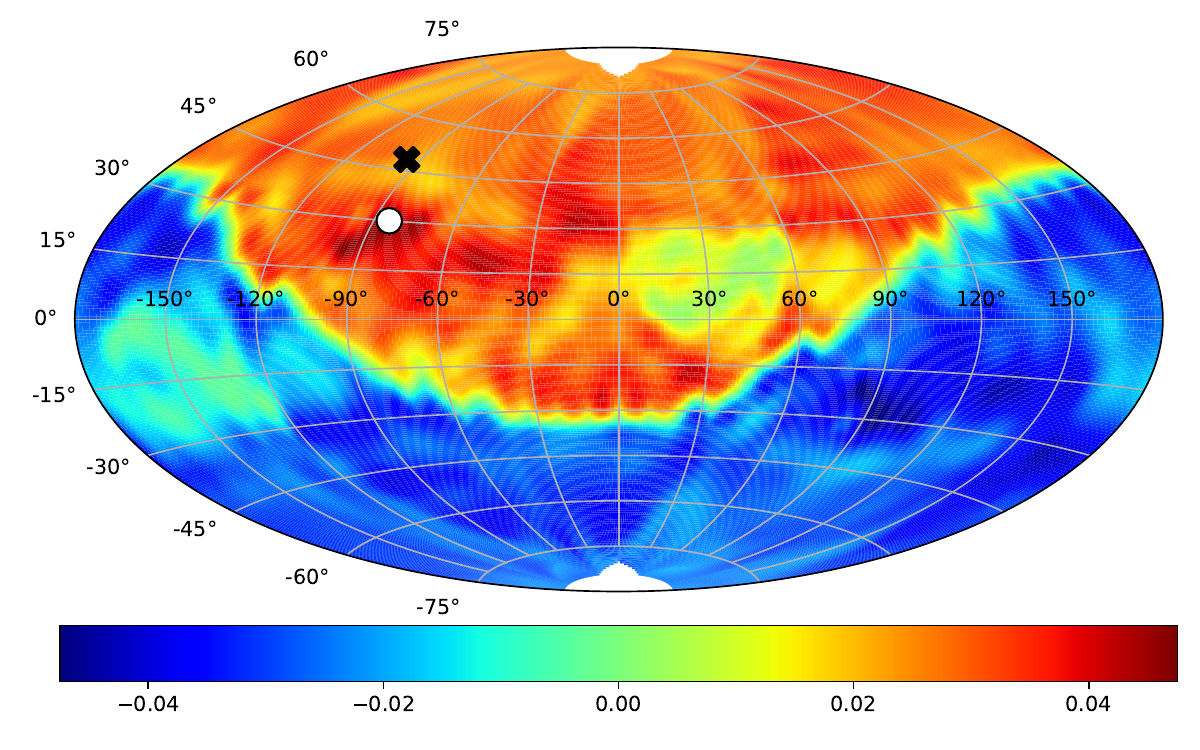}}
    \subfloat[$0.03 \leq z \leq 0.5$\label{4b}]{\includegraphics[width=0.25\textwidth,keepaspectratio]{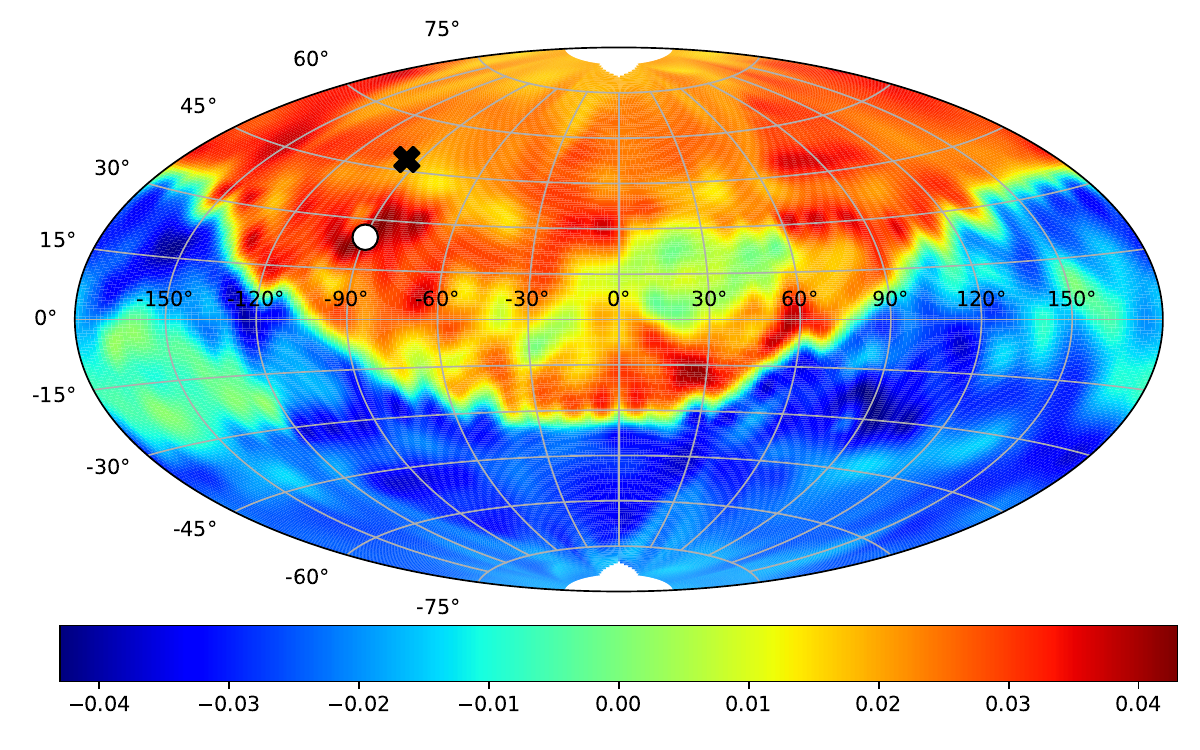}}
    \subfloat[$0.1 \leq z \leq 0.5$\label{4c}]{\includegraphics[width=0.25\textwidth,keepaspectratio]{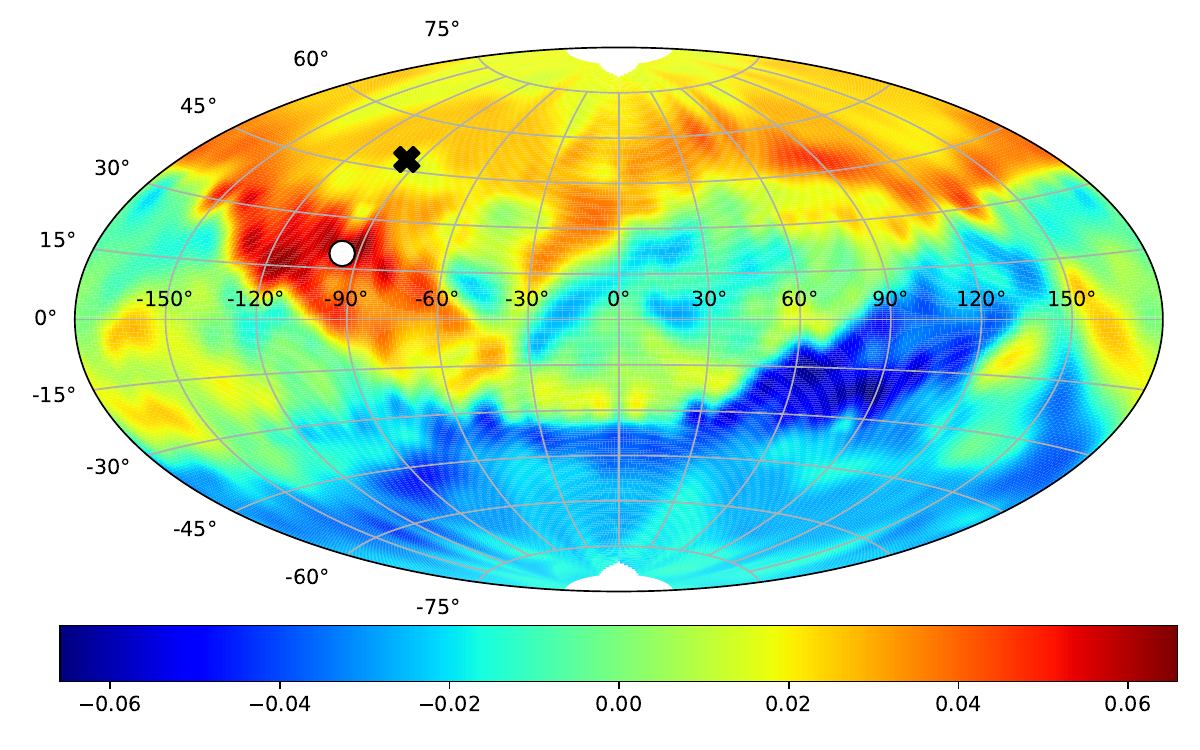}}
    \subfloat[$0.2 \leq z \leq 0.5$\label{4d}]{\includegraphics[width=0.25\textwidth,keepaspectratio]{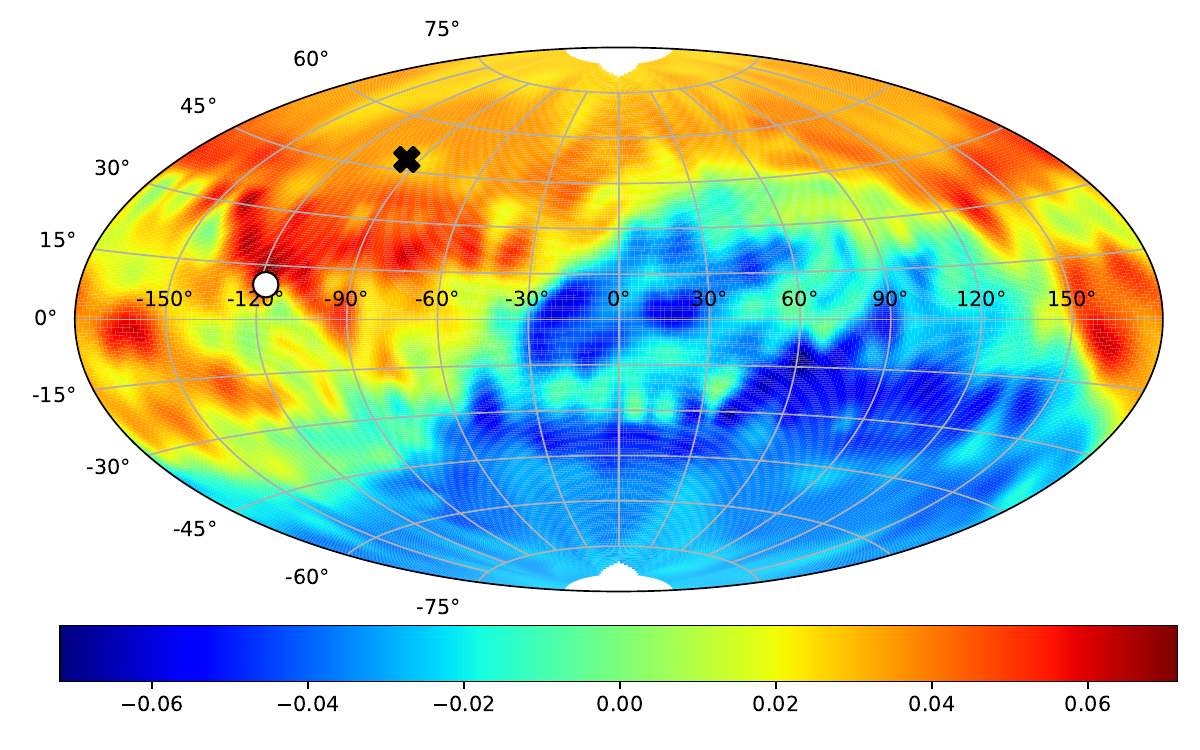}}\\
    \subfloat[$0.015 \leq z \leq 0.5$\label{4e}]{\includegraphics[width=0.25\textwidth,keepaspectratio]{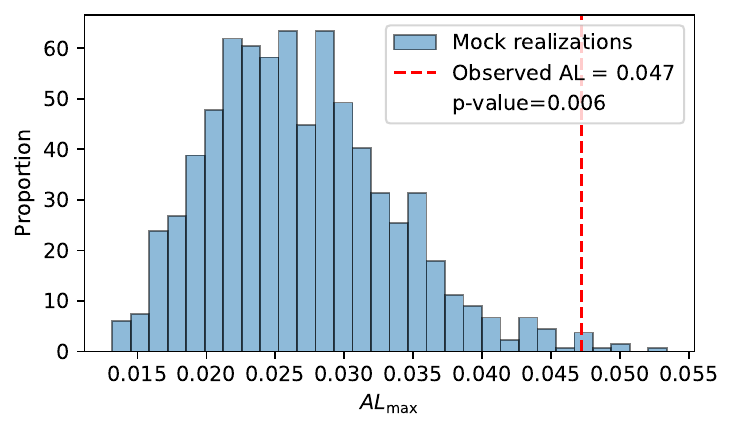}}
    \subfloat[$0.03 \leq z \leq 0.5$\label{4f}]{\includegraphics[width=0.25\textwidth,keepaspectratio]{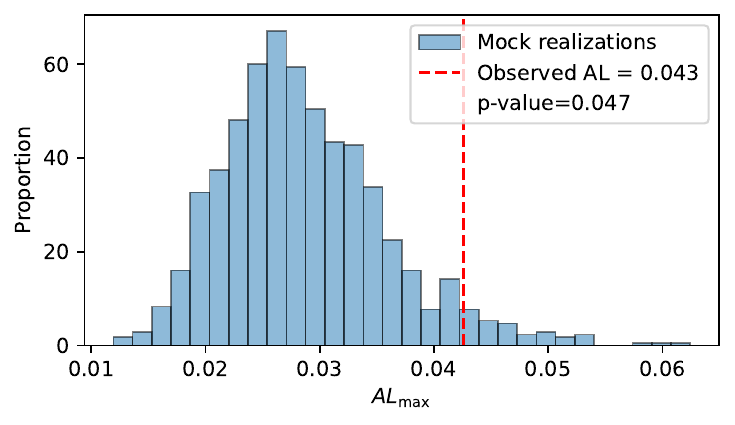}}
    \subfloat[$0.1 \leq z \leq 0.5$\label{4g}]{\includegraphics[width=0.25\textwidth,keepaspectratio]{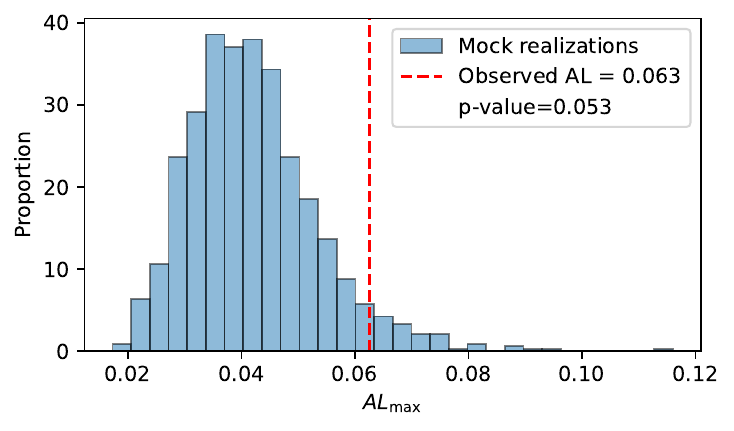}}
    \subfloat[$0.2 \leq z \leq 0.5$\label{4h}]{\includegraphics[width=0.25\textwidth,keepaspectratio]{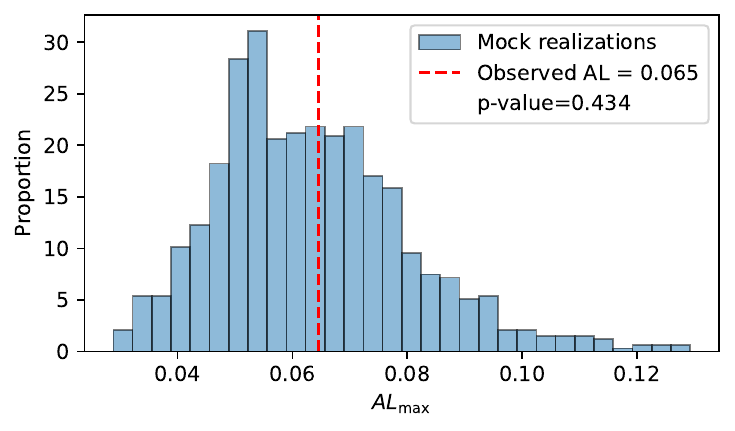}}
    \caption{Sky maps of $AL$ (first row) and the corresponding histograms of the maximum $AL$ values obtained from mock realizations (second row) for four redshift cuts (as labeled) assuming $\Lambda$CDM. The black cross marks the direction of the CMB dipole while the white circle indicates the location of maximum $AL$. This is for the SNe dataset calibrated using Cepheid-host SNe (Table~\ref{table4}). The colors correspond to the value of $AL$.}
    \label{fig4}
\end{figure}

The results for the SNe-based anisotropy analyses calibrated with using  Cepheid-host SNe are presented  in Table~\ref{table4} and Fig.~\ref{fig4}. As mentioned in Section~\ref{sec3.2}, we consider five different redshift cuts to assess the sensitivity of the inferred anisotropy to the redshift range of the SNe sample. In particular, low redshift cuts allow us to examine the impact of nearby SNe, for which local effects may be more significant. We notice that for both $\Lambda$CDM and Pad\'e-(2,1) cosmography, all the results are similar, indicating no dependence on the underlying cosmological model. The central $AL$ values increase as one progressively excludes low-redshift SNe.

Examining $\mathrm{p}_\mathrm{full}$, an anisotropy signal of $\sim 2.5 \sigma (\mathrm{p}_\mathrm{full} = 0.006)$ is observed for the redshift bin $[0.015, 0.5]$. This anisotropy signal weakens as one removes low-$z$ SNe. The maximum $AL$ direction recovered across the redshift cuts shift from $(275.62^\circ,30^\circ)$ in the $[0.015, 0.5]$ bin to $(241.88^\circ,9.59^\circ)$ in the case of the redshift bin $[0.2,0.5]$, corresponding to an angular shift of $38^\circ$. In all cases the directions are broadly consistent with, though not tightly aligned to, the CMB dipole direction, $(264.14^\circ,48.24^\circ)$, with the angular separation ranging between $\sim(21^\circ-43^\circ$).

To assess whether any observed anisotropy could be associated with a hemispherical variation in the Cepheid-host SNe calibration, we divided the 77 Cepheid-hosts into $+$ and $-$ hemispheres for seven directions (Figure~\ref{fig7}). These seven directions are the distinct maximum-$AL$ directions appearing in Tables~\ref{table4} and~\ref{table6} when Cepheid-host calibration is used: $(275.62^\circ,30^\circ)$ and $(270^\circ,24.62^\circ)$, appearing in both SNe-only and SNe$+$GC results; $(264.38^\circ,19.47^\circ)$ and $(241.88^\circ,9.59^\circ)$, appearing in SNe-only analyses; $(129.38^\circ,-9.59^\circ)$, $(157.50^\circ,-4.78^\circ)$ and $(151.88^\circ,9.59^\circ)$, found in the SNe$+$GC analysis. $M_B$ is then independently fitted in each hemisphere using the data vector $Q = m - M_B - \mu_\mathrm{ceph}$ (Eqn.~\ref{eqn6} and~\ref{eqn7}). The resulting differences in $M_B$ between the two hemispheres are negligible for all cases ($< 1.05\sigma$), providing no evidence for a statistically significant hemispherical variation in the Cepheid-host calibration that could account for the observed $AL$ values.

We carried out an additional test in which only the sky positions of the Cepheid-host SNe were shuffled, while keeping all other SNe fixed (1000 mock realizations). This test was designed to examine the role of the spatial distribution of the 77 Cepheid-host SNe on the anisotropy estimator $AL$. The $p$-value obtained from this test is denoted as p$_\mathrm{ceph}$. From Table~\ref{table4}, we find that the resulting $\mathrm{p}_\mathrm{ceph}$ values are comparable to, or smaller than, the corresponding $\mathrm{p}_\mathrm{full}$ values for the samples with $z_\mathrm{min} \leq0.1$, while both tests yield non-significant $p$-values for $z_\mathrm{min}=0.2$. This indicates that the observed $AL$ has comparable statistical significance under the two shuffling schemes. We also compute $f_\mathrm{match}$, the fraction of mocks from the Cepheid-host shuffling test for which the maximum $AL$ falls in the same \texttt{HEALPix} pixel as the maximum $AL$ obtained from the real data. $f_\mathrm{match}$ remains small ($\leq 2.9\%$) across all redshift cuts and cosmological models, indicating that random redistribution of the Cepheid-host positions rarely reproduces the exact \texttt{HEALPix} pixel corresponding to the observed maximum. Overall, this test provides no clear evidence that the observed anisotropy is primarily determined by the spatial arrangement of the Cepheid-host SNe. Combined with the absence of a statistically significant hemispherical difference in $M_B$, these results provide no clear evidence that any observed anisotropy is driven primarily by the Cepheid-host sample. Nevertheless, given the small number of Cepheid-hosts, a contribution from their spatial distribution to the low-redshift anisotropy signal cannot be excluded.

\begin{table}[ht]
\caption{Maximum $AL(H_0)$(Eq.\ref{eqn13}), its uncertainty $\delta_{AL}^{H_0}$ (Eq.\ref{eqn14}), the galactic coordinates $(l,b)$ corresponding to the maximum $AL(H_0)$ and the anisotropy $p$-values ($\mathrm{p}_\mathrm{full}$) derived from mock realizations for the SNe dataset considering five redshift cuts calibrated using Cepheid-hosts. The p$_\mathrm{ceph}$ column gives the $p$-values from a separate set of 1000 mock realizations in which only the sky positions of the Cepheid-host SNe are randomly redistributed among themselves. The $f_\mathrm{match}$ column reports the percentage of those mocks whose maximum $AL$ pixel exactly coincides with the real-data maximum $AL$ pixel.}
\label{table4}
\centering
    \begin{tabular}{c@{\hspace{1cm}}c@{\hspace{1cm}}c@{\hspace{1cm}}c@{\hspace{1cm}}c@{\hspace{1cm}}c@{\hspace{1cm}}c}
    \hline
    \thead{Redshift Cut} & \thead{$AL(H_0)$} & \thead{$\delta_{AL}^{H_0}$} & \thead{$(l,b)$} & \thead{$\mathrm{p}_\mathrm{full}$} & \thead{$\mathrm{p}_\mathrm{ceph}$} & \thead{$f_\mathrm{match}$ [$\%$]}\\
    \hline
    \hline
    & & & & & & \\[0.5ex]
    \textbf{$\Lambda$CDM} & & & & & & \\[0.5ex]
    & & & & & & \\[0.5ex]
    \textbf{$0.015 \leq z \leq 0.5$} & $0.047$ & $0.028$ & $(275.62^\circ,30^\circ)$ & $0.006$ & $0.007$ & $0.4$ \\[1.5ex]
    \textbf{$0.03 \leq z \leq 0.5$} & $0.043$ & $0.028$ & $(270^\circ,24.62^\circ)$ & $0.047$ & $0.011$ & $1.7$ \\[1.5ex]
    \textbf{$0.05 \leq z \leq 0.5$} & $0.046$ & $0.029$ & $(270^\circ,24.62^\circ)$ & $0.139$ & $0.02$ & $0.6$ \\[1.5ex]
    \textbf{$0.1 \leq z \leq 0.5$} & $0.063$ & $0.032$ & $(264.38^\circ,19.47^\circ)$ & $0.053$ & $0.026$ & $2.9$ \\[1.5ex]
    \textbf{$0.2 \leq z \leq 0.5$} & $0.065$ & $0.039$ & $(241.88^\circ,9.59^\circ)$ & $0.434$ & $0.957$ & $0.1$ \\[1.5ex]
    & & & & & & \\[0.5ex]
    \hline
    & & & & & & \\[0.5ex]
    \textbf{Pad\'e-(2,1) Cosmography} & & & & & & \\[0.5ex]
    & & & & & & \\[0.5ex]
    \textbf{$0.015 \leq z \leq 0.5$} & $0.047$ & $0.028$ & $(275.62^\circ,30^\circ)$ & $0.004$ & $0.006$ & $0.4$ \\[1.5ex]
    \textbf{$0.03 \leq z \leq 0.5$} & $0.043$ & $0.028$ & $(270^\circ,24.62^\circ)$ & $0.043$ & $0.01$ & $1.7$ \\[1.5ex]
    \textbf{$0.05 \leq z \leq 0.5$} & $0.046$ & $0.029$ & $(270^\circ,24.62^\circ)$ & $0.11$ & $0.019$ & $0.6$ \\[1.5ex]
    \textbf{$0.1 \leq z \leq 0.5$} & $0.062$ & $0.032$ & $(264.38^\circ,19.47^\circ)$ & $0.053$ & $0.025$ & $2.4$ \\[1.5ex]
    \textbf{$0.2 \leq z \leq 0.5$} & $0.064$ & $0.039$ & $(241.88^\circ,9.59^\circ)$ & $0.43$ & $0.934$ & $0.1$ \\[1.5ex]
    & & & & & & \\[0.5ex]
    \hline
    \end{tabular}
\end{table}

\begin{figure}[ht]
    \centering
    \subfloat[$0.015 \leq z \leq 0.5$\label{5a}]{\includegraphics[width=0.25\textwidth,keepaspectratio]{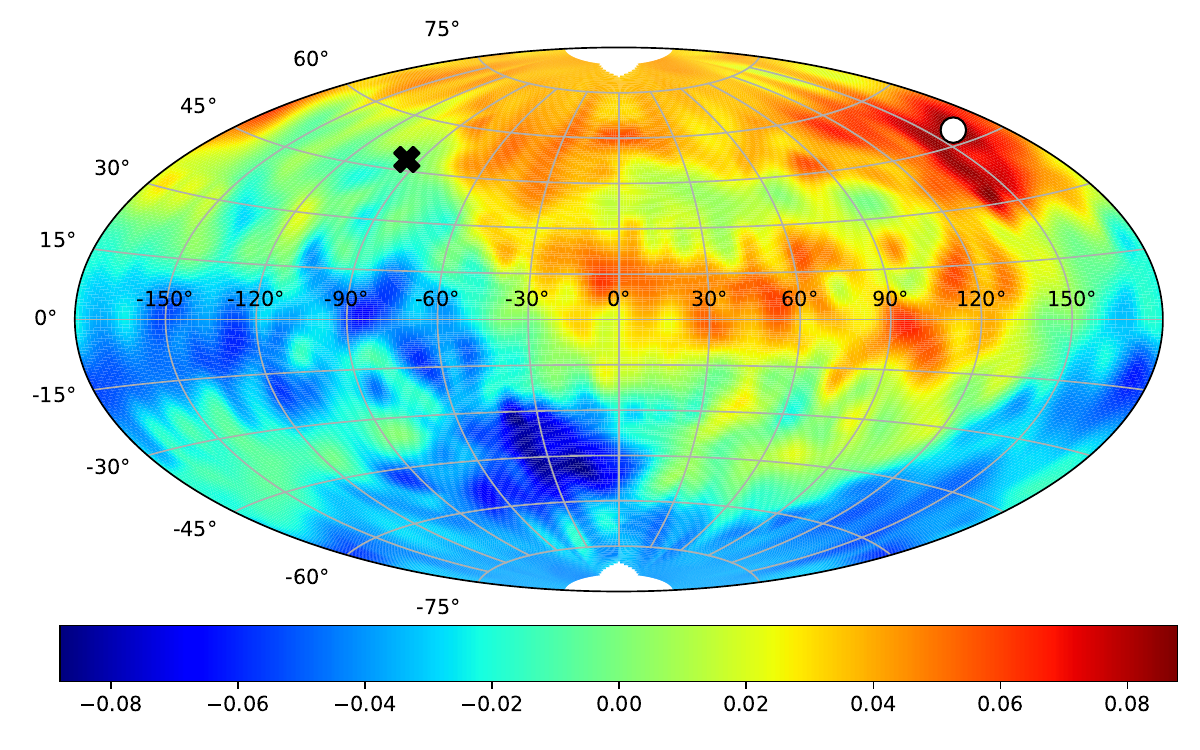}}
    \subfloat[$0.03 \leq z \leq 0.5$\label{5b}]{\includegraphics[width=0.25\textwidth,keepaspectratio]{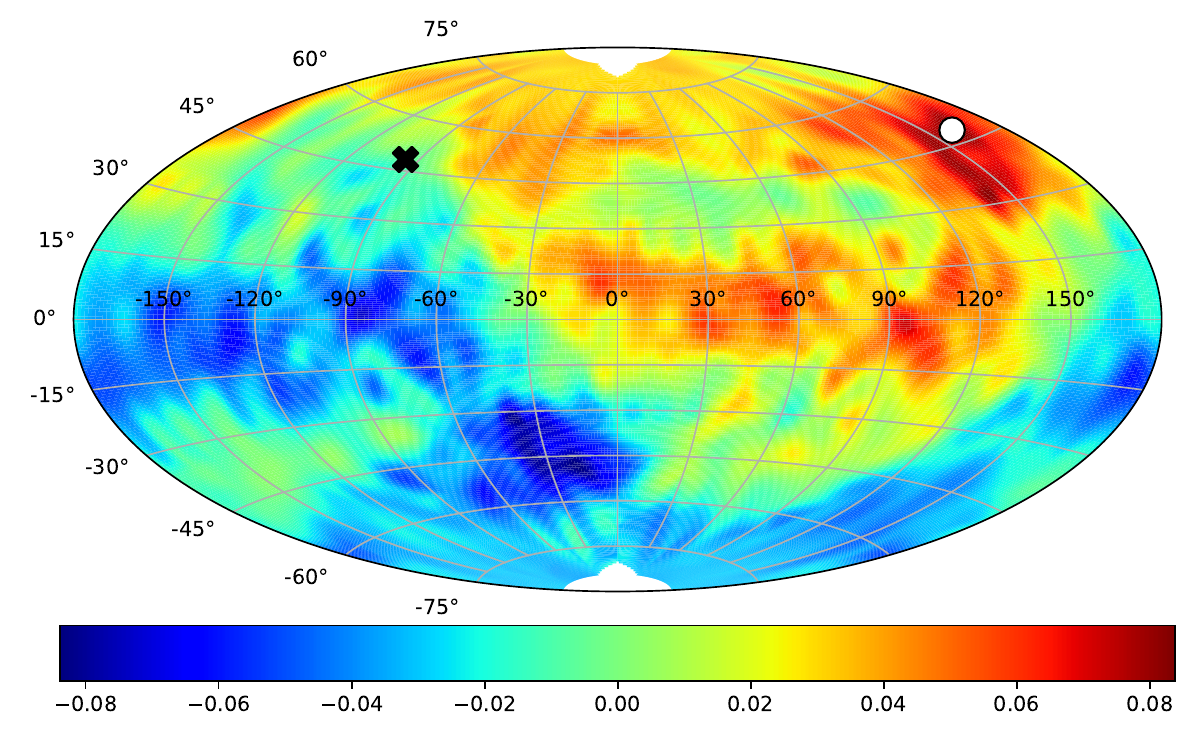}}
    \subfloat[$0.1 \leq z \leq 0.5$\label{5c}]{\includegraphics[width=0.25\textwidth,keepaspectratio]{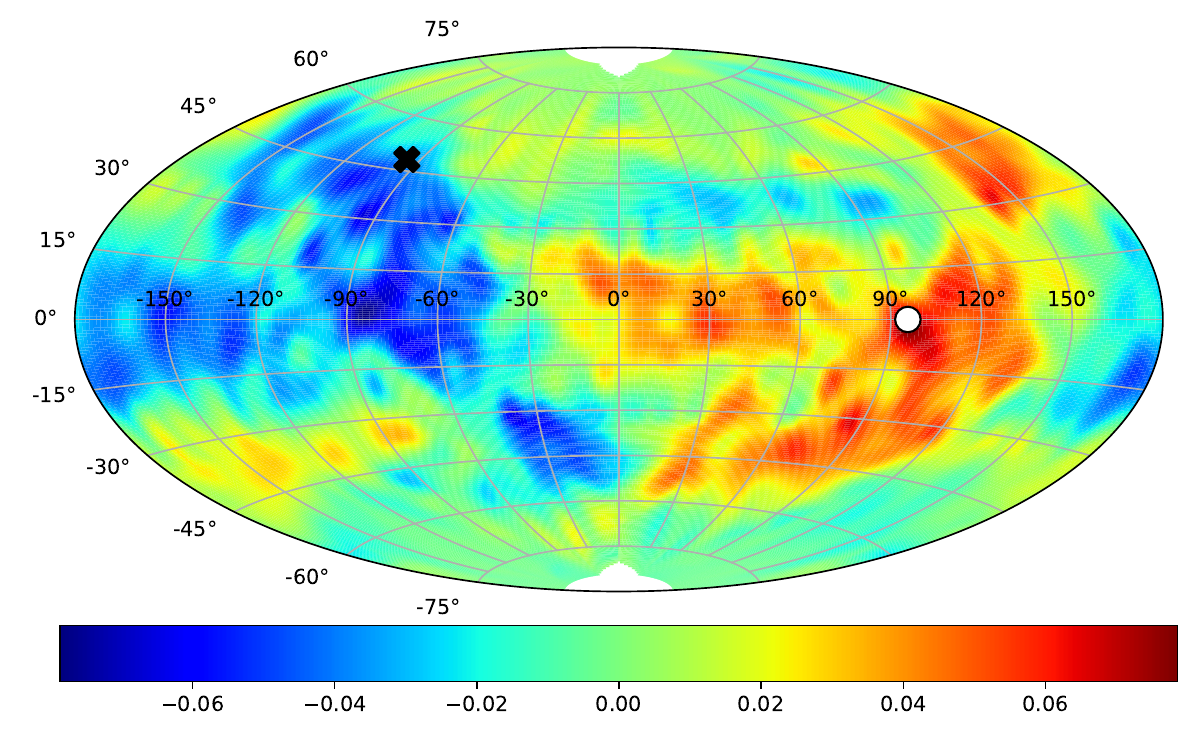}}
    \subfloat[$0.2 \leq z \leq 0.5$\label{5d}]{\includegraphics[width=0.25\textwidth,keepaspectratio]{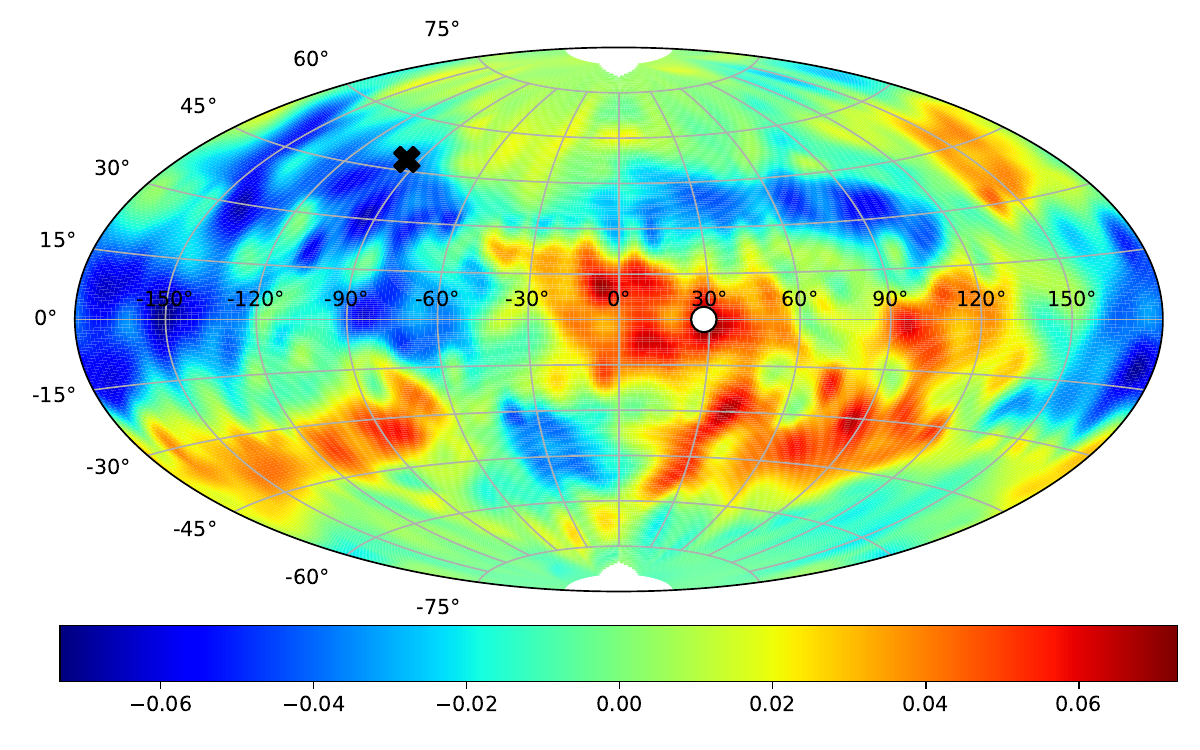}}\\
    \subfloat[$0.015 \leq z \leq 0.5$\label{5e}]{\includegraphics[width=0.25\textwidth,keepaspectratio]{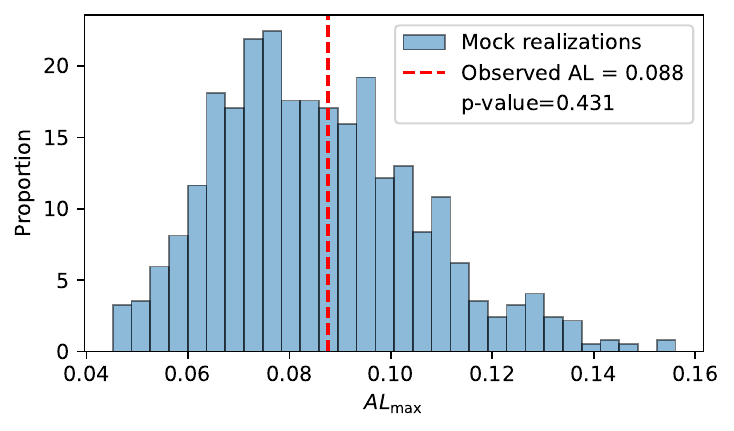}}
    \subfloat[$0.03 \leq z \leq 0.5$\label{5f}]{\includegraphics[width=0.25\textwidth,keepaspectratio]{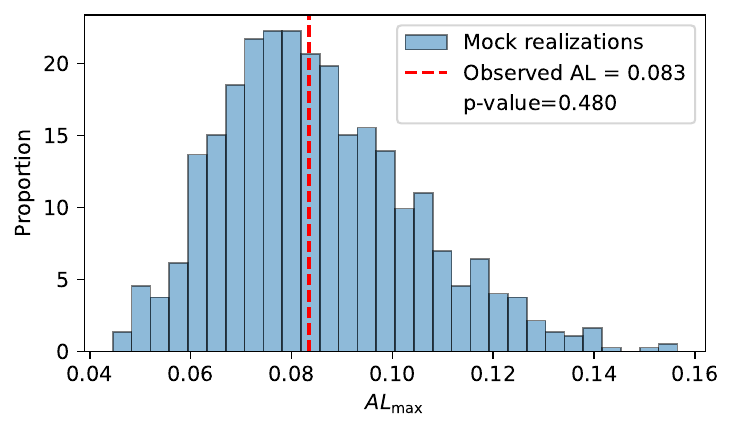}}
    \subfloat[$0.1 \leq z \leq 0.5$\label{5g}]{\includegraphics[width=0.25\textwidth,keepaspectratio]{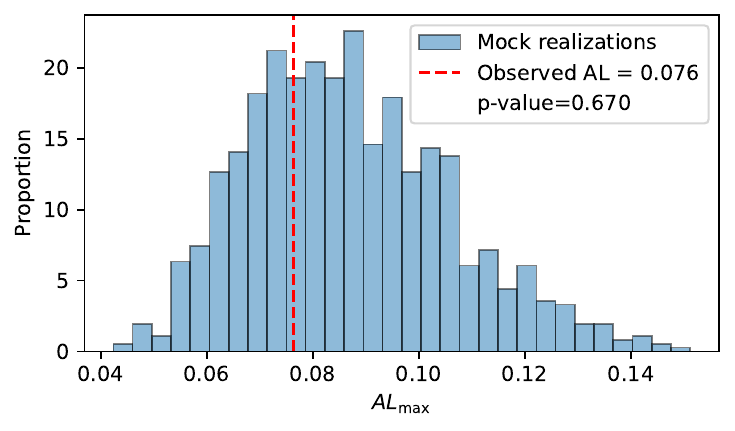}}
    \subfloat[$0.2 \leq z \leq 0.5$\label{5h}]{\includegraphics[width=0.25\textwidth,keepaspectratio]{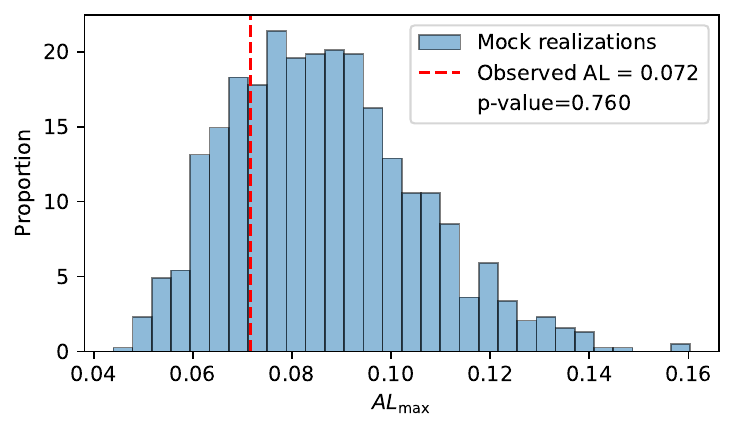}}
    \caption{Sky maps of $AL$ (first row) and the corresponding histograms of the maximum $AL$ values obtained from mock realizations (second row) for four redshift cuts (as labeled) assuming $\Lambda$CDM. The black cross marks the direction of the CMB dipole while the white circle indicates the location of maximum $AL$. This is for the SNe$+$GC dataset when $k$ is fixed to 1.14 (Table~\ref{table5}). colors correspond to the value of $AL$.}
    \label{fig5}
\end{figure}

\begin{figure}[ht]
    \centering
    \subfloat[$0.015 \leq z \leq 0.5$\label{6a}]{\includegraphics[width=0.25\textwidth,keepaspectratio]{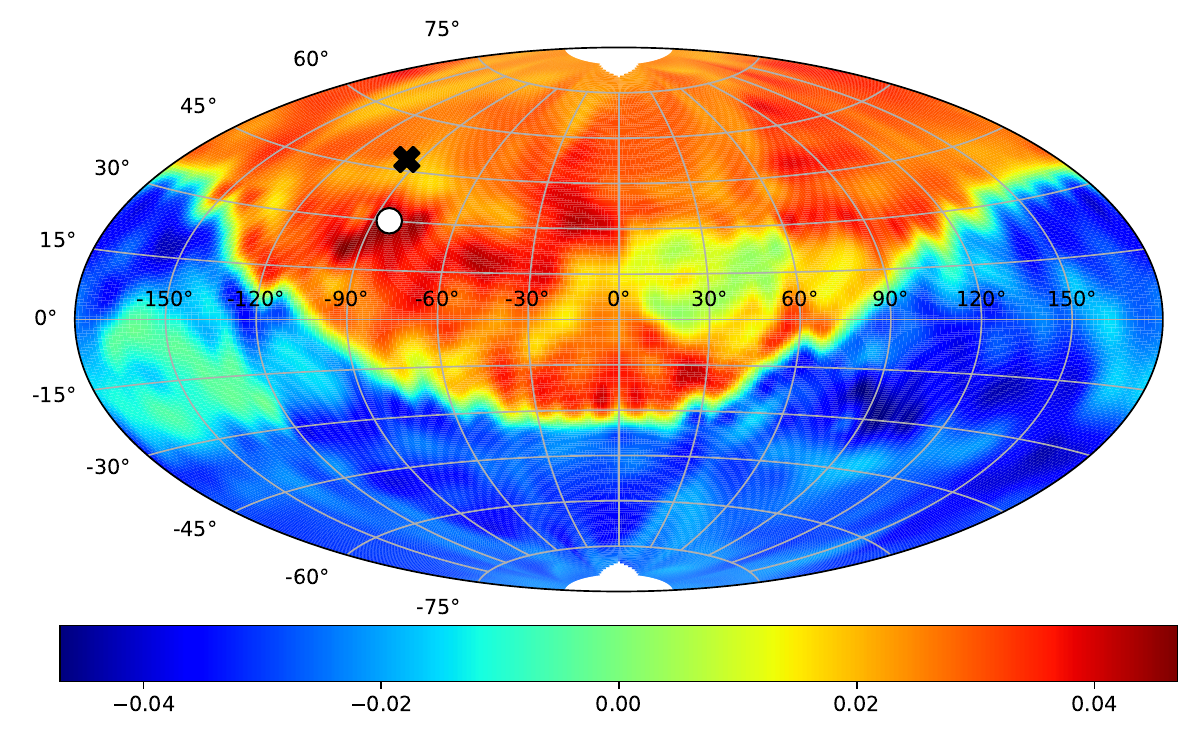}}
    \subfloat[$0.03 \leq z \leq 0.5$\label{6b}]{\includegraphics[width=0.25\textwidth,keepaspectratio]{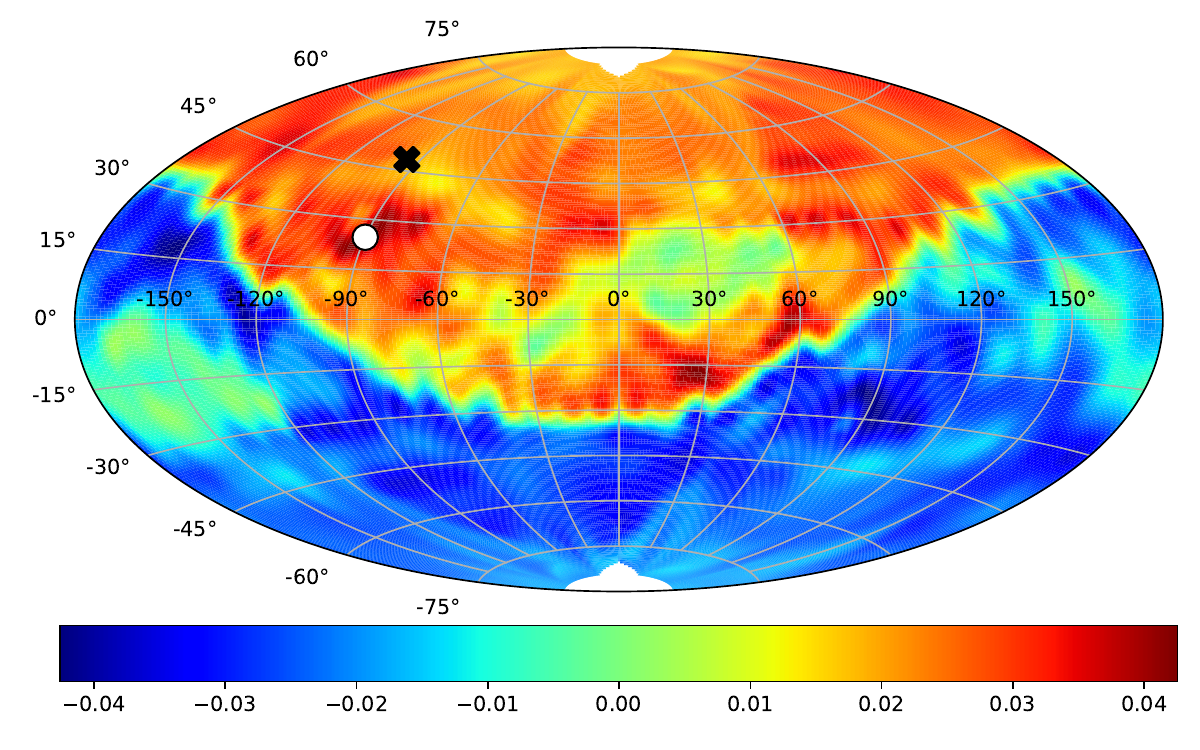}}
    \subfloat[$0.1 \leq z \leq 0.5$\label{6c}]{\includegraphics[width=0.25\textwidth,keepaspectratio]{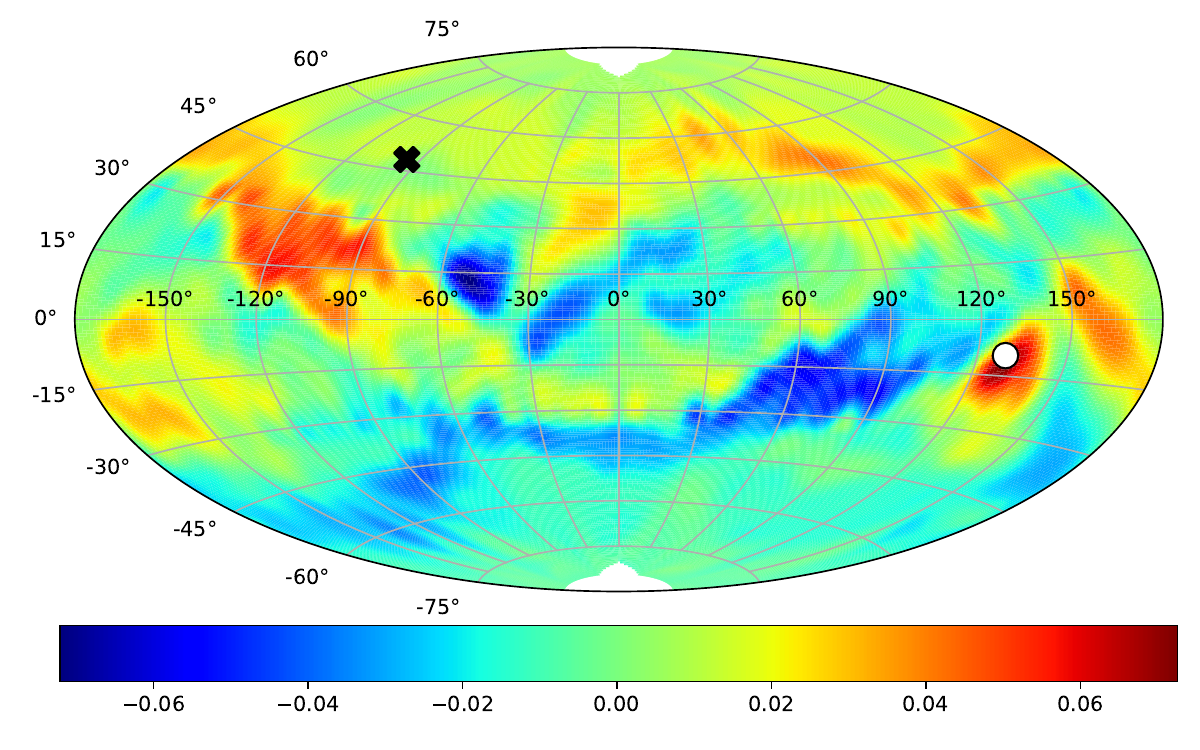}}
    \subfloat[$0.2 \leq z \leq 0.5$\label{6d}]{\includegraphics[width=0.25\textwidth,keepaspectratio]{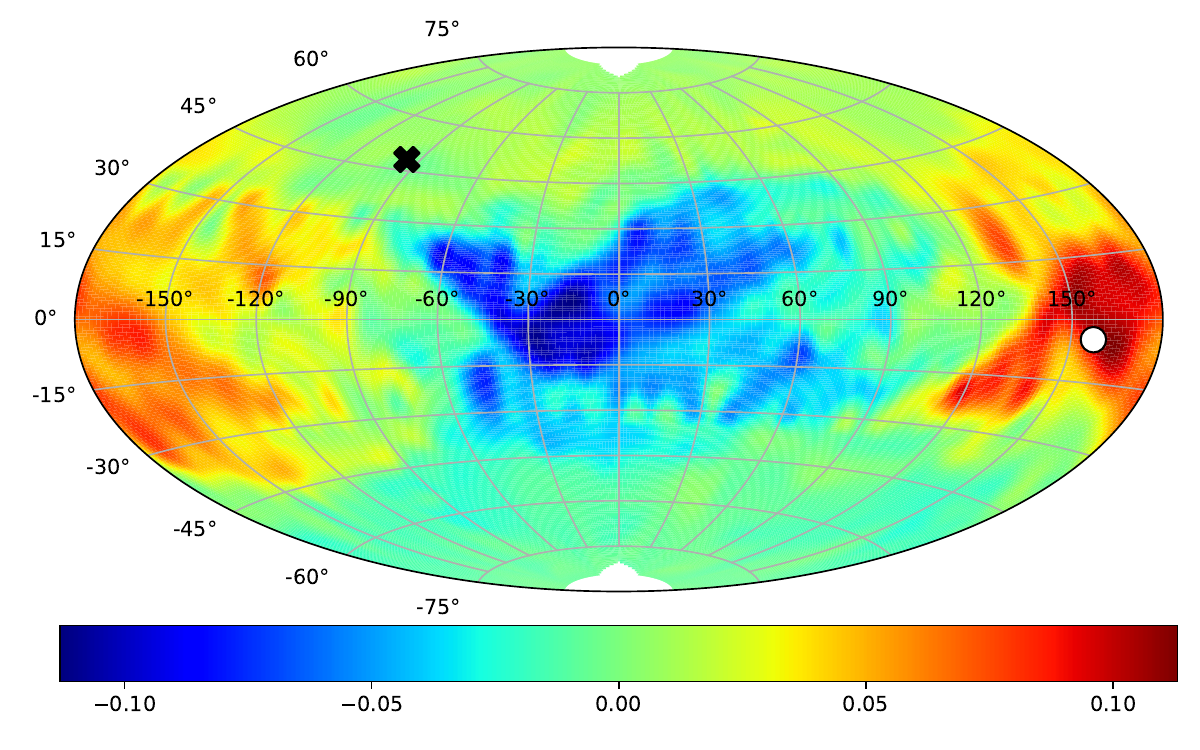}}\\
    \subfloat[$0.015 \leq z \leq 0.5$\label{6e}]{\includegraphics[width=0.25\textwidth,keepaspectratio]{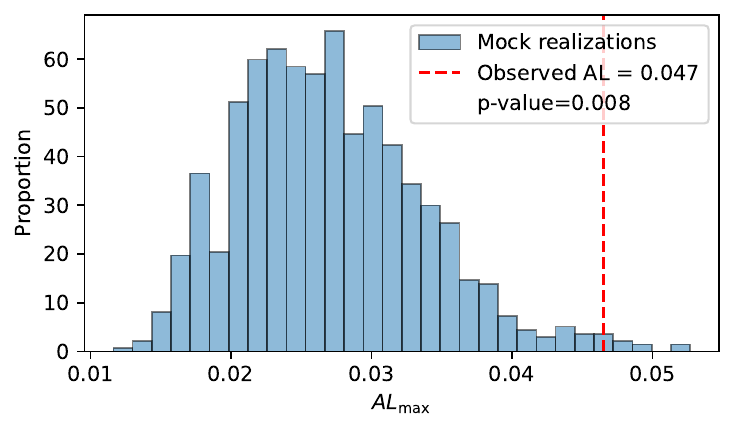}}
    \subfloat[$0.03 \leq z \leq 0.5$\label{6f}]{\includegraphics[width=0.25\textwidth,keepaspectratio]{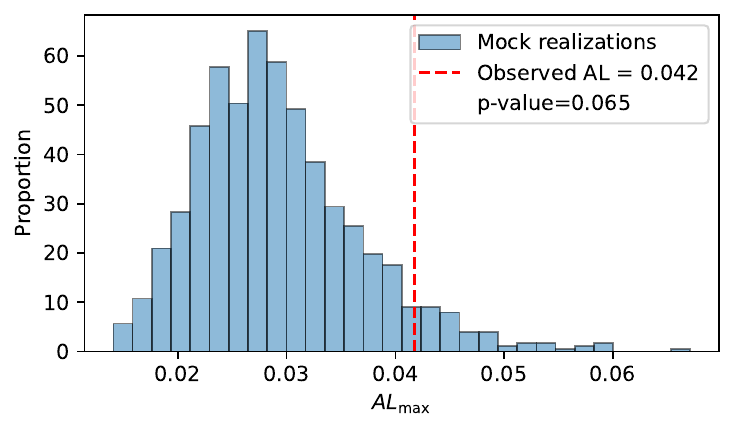}}
    \subfloat[$0.1 \leq z \leq 0.5$\label{6g}]{\includegraphics[width=0.25\textwidth,keepaspectratio]{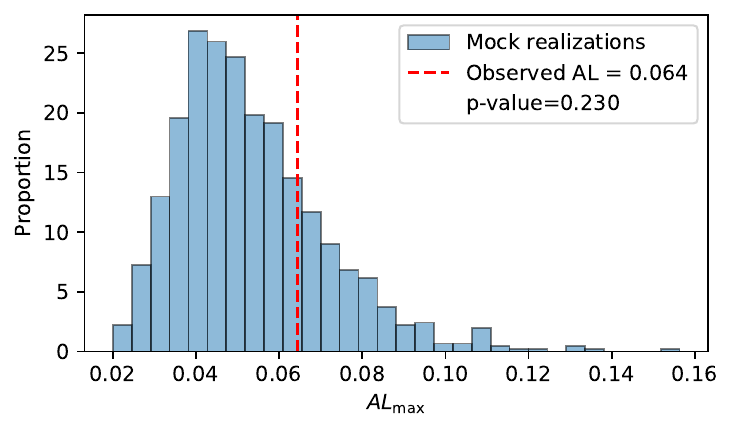}}
    \subfloat[$0.2 \leq z \leq 0.5$\label{6h}]{\includegraphics[width=0.25\textwidth,keepaspectratio]{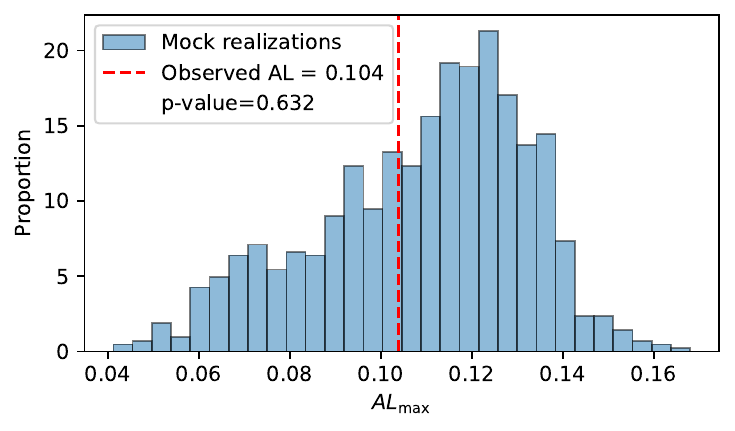}}
    \caption{Sky maps of $AL$ (first row) and the corresponding histograms of the maximum $AL$ values obtained from mock realizations (second row) for four redshift cuts (as labeled) assuming $\Lambda$CDM. The black cross marks the direction of the CMB dipole while the white circle indicates the location of maximum $AL$. This is for the SNe$+$GC dataset when Cepheid-hosts are used for calibration (Table~\ref{table6}). The colors correspond to the value of $AL$.}
    \label{fig6}
\end{figure}

\begin{figure}[ht]
    \centering
    {\includegraphics[width=0.8\textwidth,keepaspectratio]{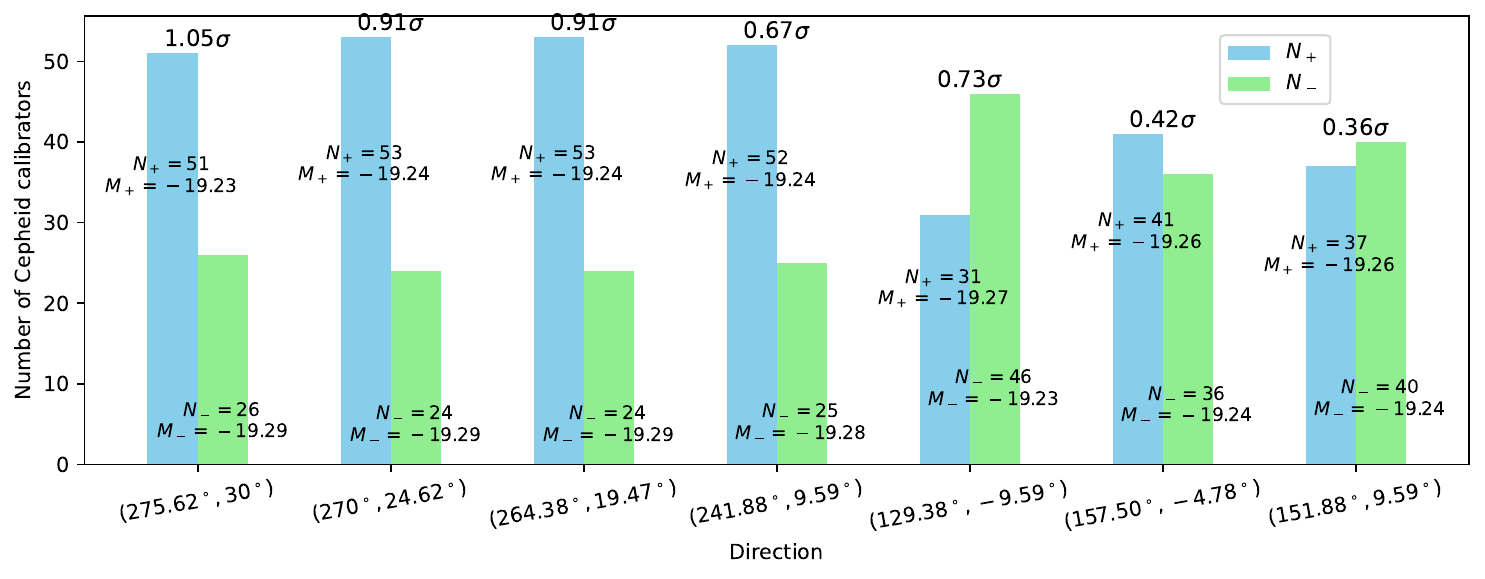}}
    \caption{The Cepheid-host SNe counts in the $+$ and $-$ hemispheres and the median $M_B$ values (errors not shown) for the seven unique directions from Tables~\ref{table4} and~\ref{table6}. The $\sigma$ values indicate the significance in the difference between $M_+$ and $M_-$ where $M_+$ and $M_-$ refer to the values of $M_B$ in the $+$ and $-$ hemispheres. }
    \label{fig7}
\end{figure}

The results for SNe$+$GC combination are presented in Tables~\ref{table5} (fixed $k$) and~\ref{table6} (Cepheid-host calibrated) and Figs.~\ref{fig5} and~\ref{fig6}. This part of the analysis is intended to assess the robustness of the anisotropy preference found in the SNe-only sample by incorporating GC data as a complementary probe of anisotropy through their X-ray scaling relations. In particular, we examine whether the anisotropy preference observed in the SNe-only analysis persists when GC data are included. For all SNe redshift cuts, the full GC sample of 313 clusters is retained. This choice allows us to isolate the effect of progressively removing low-redshift SNe, without introducing additional variations arising from changes in the GC sample.

As in the GC-only and SNe-only analyses, the inferred anisotropy is largely insensitive to the choice of the cosmological model. The maximum $AL$ directions and $p$-values obtained under $\Lambda$CDM and Pad\'e-(2,1) cosmography are generally very similar. In contrast, the maximum $AL$ direction is sensitive to the treatment adopted for the $H_0-M_B$ and $H_0-k$ degeneracies. The angular separation between the maximum $AL$ directions obtained using the fixed $k$ and Cepheid-host calibrated approaches range from $80^\circ-150^\circ$ for matching redshift cuts and cosmological models. This suggests that the treatment of the $H_0-M_B$ and $H_0-k$ degeneracies have an impact on the maximum $AL$ direction rather than the choice of cosmological models considered in this work. However, since all $p$-values in the fixed k analysis (Table~\ref{table5}) are consistent with statistical isotropy, these directional variations should not be interpreted as evidence for a physical anisotropy. Instead they suggest that when the anisotropy significance is weak, the location of the maximum $AL$ direction may become sensitive to statistical fluctuations and details of the analysis methodology.

From Table~\ref{table6}, when using the Cepheid-host calibration, we find that the lowest-redshift cut, $z_\mathrm{min}=0.015$, shows a mild anisotropy preference with an $AL$ significance of $\sim 2.4 \sigma$ ($\mathrm{p}_\mathrm{full}=0.008$). This preference weakens as the lowest-redshift SNe are progressively removed, with $\mathrm{p}_\mathrm{full} = 0.065$ and $0.262$ for $z_\mathrm{min}=0.03$ and $0.05$. Furthermore, the maximum $AL$ directions for these three redshift cuts are similar to those obtained in the SNe-only analysis. The contrast with the fixed-$k$ results (Table~\ref{table5}), which are consistent with statistical isotropy for all redshift cuts, indicates that the inferred anisotropy preference depends on the treatment adopted to constrain $H_0$, namely, whether $k$ is fixed or the calibration is anchored by the Cepheid-host SNe.

\begin{table}[ht]
\caption{Maximum $AL(H_0)$(Eq.\ref{eqn13}), its uncertainty $\delta_{AL}^{H_0}$ (Eq.\ref{eqn14}), the galactic coordinates $(l,b)$ corresponding to the maximum $AL(H_0)$ and the anisotropy $p$-values ($\mathrm{p}_\mathrm{full}$) derived from mock realizations for the SNe$+$combined GC dataset combination considering five redshift cuts (on the SNe sample) while fixing $k$.}
\label{table5}
\centering
    \begin{tabular}{c@{\hspace{1cm}}c@{\hspace{1cm}}c@{\hspace{1cm}}c@{\hspace{1cm}}c}
    \hline
    \thead{Redshift Cut} & \thead{$AL(H_0)$} & \thead{$\delta_{AL}^{H_0}$} & \thead{$(l,b)$} & \thead{$\mathrm{p}_\mathrm{full}$}\\
    \hline
    \hline
    & & & & \\[0.5ex]
    \textbf{$\Lambda$CDM} & & & & \\[0.5ex]
    & & & & \\[0.5ex]
    \textbf{$0.015 \leq z \leq 0.5$} & $0.088$ & $0.034$ & $(160.71^\circ,48.14^\circ)$ & $0.431$ \\[1.5ex]
    \textbf{$0.03 \leq z \leq 0.5$} & $0.084$ & $0.034$ & $(160.71^\circ,48.14^\circ)$ & $0.480$ \\[1.5ex]
    \textbf{$0.05 \leq z \leq 0.5$} & $0.076$ & $0.034$ & $(95.62^\circ,0^\circ)$ & $0.680$ \\[1.5ex]
    \textbf{$0.1 \leq z \leq 0.5$} & $0.076$ & $0.034$ & $(95.62^\circ,0^\circ)$ & $0.670$ \\[1.5ex]
    \textbf{$0.2 \leq z \leq 0.5$} & $0.072$ & $0.036$ & $(28.12^\circ,0^\circ)$ & $0.760$ \\[1.5ex]
    & & & & \\[0.5ex]
    \hline
    & & & & \\[0.5ex]
    \textbf{Pad\'e-(2,1) Cosmography} & & & & \\[0.5ex]
    & & & & \\[0.5ex]
    \textbf{$0.015 \leq z \leq 0.5$} & $0.088$ & $0.034$ & $(160.71^\circ,48.14^\circ)$ & $0.429$ \\[1.5ex]
    \textbf{$0.03 \leq z \leq 0.5$} & $0.084$ & $0.034$ & $(160.71^\circ,48.14^\circ)$ & $0.471$ \\[1.5ex]
    \textbf{$0.05 \leq z \leq 0.5$} & $0.075$ & $0.034$ & $(95.62^\circ,0^\circ)$ & $0.688$ \\[1.5ex]
    \textbf{$0.1 \leq z \leq 0.5$} & $0.076$ & $0.034$ & $(95.62^\circ,0^\circ)$ & $0.678$ \\[1.5ex]
    \textbf{$0.2 \leq z \leq 0.5$} & $0.074$ & $0.034$ & $(270^\circ,-35.69^\circ)$ & $0.720$ \\[1.5ex]
    & & & & \\[0.5ex]
    \hline
    \end{tabular}
\end{table}

\begin{table}[ht]
\caption{Maximum $AL(H_0)$(Eq.\ref{eqn13}), its uncertainty $\delta_{AL}^{H_0}$ (Eq.\ref{eqn14}), the galactic coordinates $(l,b)$ corresponding to the maximum $AL(H_0)$ and the anisotropy $p$-values ($\mathrm{p}_\mathrm{full}$) derived from mock realizations for the SNe$+$combined GC dataset combination considering five redshift cuts (on the SNe sample) calibrated using Cepheid-hosts. The p$_\mathrm{ceph}$ column gives the $p$-values from a separate set of 1000 mock realizations in which only the sky positions of the Cepheid-host SNe are randomly redistributed among themselves. The $f_\mathrm{match}$ column reports the percentage of those mocks whose maximum $AL$ pixel exactly coincides with the real-data maximum $AL$ pixel.}
\label{table6}
\centering
    \begin{tabular}{c@{\hspace{1cm}}c@{\hspace{1cm}}c@{\hspace{1cm}}c@{\hspace{1cm}}c@{\hspace{1cm}}c@{\hspace{1cm}}c}
    \hline
    \thead{Redshift Cut} & \thead{$AL(H_0)$} & \thead{$\delta_{AL}^{H_0}$} & \thead{$(l,b)$} & \thead{$\mathrm{p}_\mathrm{full}$} & \thead{$\mathrm{p}_\mathrm{ceph}$} & \thead{$f_\mathrm{match}$ [$\%$]}\\
    \hline
    \hline
    & & & & & & \\[0.5ex]
    \textbf{$\Lambda$CDM} & & & & & & \\[0.5ex]
    & & & & \\[0.5ex]
    \textbf{$0.015 \leq z \leq 0.5$} & $0.047$ & $0.028$ & $(275.62^\circ,30^\circ)$ & $0.008$ & $0.007$ & $0.3$ \\[1.5ex]
    \textbf{$0.03 \leq z \leq 0.5$} & $0.042$ & $0.028$ & $(270^\circ,24.62^\circ)$ & $0.065$ & $0.012$ & $1.6$ \\[1.5ex]
    \textbf{$0.05 \leq z \leq 0.5$} & $0.044$ & $0.030$ & $(270^\circ,24.62^\circ)$ & $0.262$ & $0.031$ & $0.5$ \\[1.5ex]
    \textbf{$0.1 \leq z \leq 0.5$} & $0.065$ & $0.040$ & $(129.38^\circ,-9.59^\circ)$ & $0.230$ & $0.995$ & $27.9$ \\[1.5ex]
    \textbf{$0.2 \leq z \leq 0.5$} & $0.104$ & $0.030$ & $(157.50^\circ,-4.78^\circ)$ & $0.632$ & $1.0$ & $2.4$ \\[1.5ex]
    & & & & & & \\[0.5ex]
    \hline
    & & & & & & \\[0.5ex]
    \textbf{Pad\'e-(2,1) Cosmography} & & & & & & \\[0.5ex]
    & & & & & & \\[0.5ex]
    \textbf{$0.015 \leq z \leq 0.5$} & $0.047$ & $0.028$ & $(275.62^\circ,30^\circ)$ & $0.007$ & $0.007$ & $0.3$ \\[1.5ex]
    \textbf{$0.03 \leq z \leq 0.5$} & $0.042$ & $0.028$ & $(270^\circ,24.62^\circ)$ & $0.063$ & $0.012$ & $1.6$ \\[1.5ex]
    \textbf{$0.05 \leq z \leq 0.5$} & $0.044$ & $0.029$ & $(270^\circ,24.62^\circ)$ & $0.246$ & $0.031$ & $0.5$ \\[1.5ex]
    \textbf{$0.1 \leq z \leq 0.5$} & $0.060$ & $0.039$ & $(129.38^\circ,-9.59^\circ)$ & $0.264$ & $0.996$ & $28.2$ \\[1.5ex]
    \textbf{$0.2 \leq z \leq 0.5$} & $0.116$ & $0.030$ & $(151.88^\circ,9.59^\circ)$ & $0.619$ & $0.999$ & $0.1$ \\[1.5ex]
    & & & & & & \\[0.5ex]
    \hline
    \end{tabular}
\end{table}

In Figure~\ref{fig7}, we find that the significance for the $M_B$ hemispherical difference is $\lesssim 1.05\sigma$ for the distinct directions appearing in Table~\ref{table6}, which reports results for the SNe$+$GC dataset combination calibrated using Cepheid-host SNe. We also find that $\mathrm{p}_\mathrm{ceph}$ values show a weakening of the $AL$ significance as low-$z$ SNe are removed; this is similar to the SNe-only analysis and indicates that the observed $AL$ has comparable statistical significance under the full-sky and Cepheid-host shuffling schemes. The $f_\mathrm{match}$ values are generally small, indicating that the exact maximum $AL$ is rarely reproduced under Cepheid-host shuffling, although the $z_\mathrm{min}=0.1$ case yields a substantially larger value of $28\%$. This case is nevertheless consistent with statistical isotropy, as indicated by its non-significant $\mathrm{p}_\mathrm{full}$ and $\mathrm{p}_\mathrm{ceph}$. Our conclusion from the SNe-only analysis therefore carries over to the SNe$+$GC combination: we find no clear evidence that any observed anisotropy is primarily driven by the Cepheid-host calibrators, although a contribution from their spatial distribution cannot be excluded due to the small number of Cepheid-hosts.

The combined GC-only analysis showed no robust anisotropy, while the SNe-only analysis showed only a mild low-$z$ preference. Therefore, the disappearance of the signal in the joint analysis is consistent with the absence of a strong anisotropic signal in either dataset individually.

\section{Conclusions}
\label{sec5}

In this work, we have studied anisotropy in $H_0$ using GC and SNe datasets by decomposing the sky into two hemispheres. Since SNe are commonly calibrated using Cepheid-host SNe when employed to study cosmic anisotropy, the small number of Cepheid-hosts can introduce statistical fluctuations in the inferred absolute-magnitude calibration~\cite{colgain_2023}. To investigate the robustness of any anisotropy preference in the SNe sample, we include GC, which serve as a complementary observational probe of anisotropy through their X-ray scaling relations. We consider two approaches for constraining $H_0$: Cepheid-host SNe calibration and fixing the normalization parameter $k$ of the GC $L_X-T$ scaling relation to its best-fit value obtained under a fiducial cosmology. Our results are tabulated in Tables~\ref{table3}-\ref{table6}. We consider both the standard $\Lambda$CDM model and the model-independent Pad\'e-(2,1) cosmography (with $j_0=1$) to assess the dependence of any inferred anisotropy on the assumed cosmological model.

Our findings indicate that the Chandra and the combined GC sample show consistency with statistical isotropy unlike XMM-Newton. The maximum $AL$ direction for XMM-Newton differs by $\sim 138^\circ$ from that of the combined sample. Previous studies~\cite{mikgas_2021, He_2025} have shown that $L_X-T$ relation is a relatively noisy anisotropy tracer. Taken together, this suggests that the anisotropy preference in the XMM-Newton sample may arise from statistical fluctuations associated with its small sample size, rather than providing evidence for a robust anisotropy. \rthis{This is supported by the maximum normalized statistic, max($\sigma$), which is not unusual compared with the isotropic mock realizations ($\mathrm{p}_\mathrm{\sigma} \sim 0.161$) }. We find an anisotropy preference ($\sim 2.5\sigma$) when conducting the SNe-only analysis in the redshift bin $[0.015, 0.5]$. Progressively removing low-$z$ SNe decreases the significance, suggesting that this anisotropy preference is associated with the low-redshift SNe sample. For the SNe$+$GC analysis, we find that the treatment of the $H_0-M_B$ and $H_0-k$ degeneracies affects any inferred anisotropy preference: fixing $k$ yields consistency with statistical isotropy for all redshift cuts, whereas using Cepheid-hosts to calibrate $M_B$ yields an anisotropy preference for the low-redshift cuts which weakens as low-$z$ SNe are removed.

For cases where Cepheid-host calibration is used, we performed an additional test in which only the sky positions of the 77 Cepheid-host SNe were redistributed. The resulting $p$-values ($\mathrm{p}_\mathrm{ceph}$) show a similar trend to those obtained from full data shuffling, with the anisotropy preference weakening as the low-$z$ SNe are progressively removed. This provides no clear evidence that the spatial distribution of Cepheid-host SNe is responsible for the observed anisotropy preference. We also find no statistically significant hemispherical differences ($\leq 1.05\sigma$) in $M_B$, arguing against a strong directional calibration bias in the Cepheid sample. We also compute $f_\mathrm{match}$ which indicates the fraction of mocks from the Cepheid-host shuffling test for which the maximum $AL$ falls in the same \texttt{HEALPix} pixel as the maximum $AL$ obtained from the real data. We find, in most cases, that $f_\mathrm{match}$ values are small indicating that redistribution of the Cepheid-host positions rarely reproduces the exact \texttt{HEALPix} pixel corresponding to the observed maximum. Together, these tests provide no evidence that the observed anisotropy preference is driven primarily by the Cepheid-host sample, although a modest contribution from the calibrators cannot be excluded.

Overall, our results do not provide compelling evidence for a statistically significant anisotropy in the local expansion rate. The anisotropy preference is strongest in the low-redshift SNe sample and weakens as low-$z$ SNe are progressively removed, while the inclusion of GC data yields consistency with statistical isotropy when the $L_X-T$ normalization ($k$) is fixed. The persistence of a mild preference in the Cepheid-host calibrated SNe$+$GC analysis suggests that the low-redshift preference may be sensitive to effects such as the local velocity field or calibration uncertainties~\cite{yasin_2026}, rather than providing robust evidence for a large-scale departure from isotropic expansion. While our tests do not identify a specific source for the low-redshift anisotropy preference, they indicate that neither the directional calibration of the Cepheid-hosts nor their spatial distribution provides a compelling explanation. Thus, the observed low-redshift anisotropy preference should be interpreted with caution and does not currently provide a robust detection of cosmic anisotropy.

Our findings (for SNe-only and SNe$+$GC calibrated using Cepheid-hosts) are consistent with recent studies of the PantheonPlus sample by Ref.~\cite{colgain_2026} which report a decrease in both dipole amplitude and statistical significance with increasing $z_\mathrm{min}$. This suggests that any anisotropy preference inferred from the SNe sample is primarily associated with the low-$z$ universe.

\rthis{We would like to point out that the significance of the anisotropy for the GC-only case where $k$ is fixed should be interpreted in the context of the adopted null distribution.} In our analysis, the measured cluster properties are reshuffled among the fixed sky positions of the observed sample (Table~\ref{table1}) and the significance is determined from the distribution of the maximum $AL$ over the full sky scan. This differs from the bootstrap-based approach of~\cite{mikgas_2020} and the independent \texttt{FLAMINGO} lightcone realizations used by~\cite{He_2025}, which adopt different null distributions and sampling statistics. Consequently, differences between the reported numerical significances do not necessarily indicate an inconsistency, but can arise from different assumptions and realizations underlying the isotropic null hypotheses.

The method employed in this work is sensitive only to dipolar anisotropy. A more detailed analysis involving higher-order multipoles in SNe datasets could reveal additional features that cannot be captured by hemispherical comparison alone. Future studies could also incorporate modelling of the local peculiar-velocity field and its possible anisotropy, particularly given the stronger preference at low-redshifts~\cite{yasin_2026}. Broadly, our results highlight the importance of combining multiple cosmological probes when testing anisotropy, as different datasets may be subject to distinct statistical and systematic uncertainties. Since the GC scaling relations play a central role in the present analysis, it would be interesting to repeat the study using alternative cluster observables and scaling relations. Exploring additional SNe redshift cuts could further test the stability of the results obtained in this work. Recent studies have highlighted the potential impact of SNe progenitor-age bias~\cite{Sah_2026} (and references therein), making it worthwhile to investigate cosmological anisotropy while accounting for possible progenitor-age effects. Extending the present analysis to additional independent observational probes and larger, homogeneous samples will ultimately be important for determining whether the low-redshift anisotropy preference persists and for distinguishing a genuine cosmological signal from local or observational effects.

We have made our analysis codes publicly available, which can be found at
\url{https://github.com/Perallheizer/Anisotropy.git}


\begin{acknowledgments}
SB would like to extend his gratitude to the University Grants Commission (UGC), Govt. of India for their continuous support through the Senior Research Fellowship, which has played a crucial role in the successful completion of our research. We would also like to thank Eoin \'O Colg\'ain and Pedro da Silveira Ferreira for valuable comments on this work. We are also grateful to the anonymous referee for constructive feedback and useful comments on our manuscript.
The computational work used for this analysis was supported by the National Supercomputing Mission (NSM), Government of India, through access to the ``PARAM SEVA'' facility at IIT Hyderabad. The NSM is implemented by the Centre for Development of Advanced Computing (C-DAC) with funding from the Ministry of Electronics and Information Technology (MeitY) and the Department of Science and Technology (DST). We also acknowledge the use of IUCAA HPC Computing facilities.
\end{acknowledgments}


\bibliography{references}

\end{document}